\documentclass[10pt,prb,aps,twocolumn,showpacs,floatfix]{revtex4}

\usepackage{graphicx}
\usepackage{amssymb}
\usepackage{amsmath} 
\usepackage{tensor}
 
\usepackage{color}
\newcommand{\up}{\uparrow}
\newcommand{\dn}{\downarrow}

\begin{document}

\title{Quantum Monte Carlo studies of spinons in one-dimensional spin systems}

\author{Ying Tang}
\affiliation{Department of Physics, Boston University, 590 Commonwealth Avenue, Boston, Massachusetts 02215, USA}

\author{Anders W. Sandvik}
\affiliation{Department of Physics, Boston University, 590 Commonwealth Avenue, Boston, Massachusetts 02215, USA}

\begin{abstract}
Observing constituent particles with fractional quantum numbers in confined and deconfined states is an interesting and challenging problem in quantum many-body 
physics. Here we further explore a computational scheme [Y. Tang and A. W. Sandvik, Phys.~Rev.~Lett.~{\bf 107}, 157201 (2011)] based on valence-bond quantum Monte 
Carlo simulations of quantum spin systems. Using several different one-dimensional models, we characterize $S=1/2$ spinon excitations using the intrinsic spinon size 
$\lambda$ and confinement length $\Lambda$ (the size of a bound state). The spinons have finite size in valence-bond-solid states, infinite size in the critical 
region (with overlaps characterized by power laws), and become ill defined (completely unlocalizable) in the N\'eel state (which we stabilize in one dimension by 
introducing long-range interactions). We also verify that pairs of spinons are deconfined in uniform spin chains but become confined upon introducing a 
pattern of alternating coupling strengths (dimerization) or coupling two chains (forming a ladder). In the dimerized system, an individual spinon can be small when 
the confinement length is large; this is the case when the imposed dimerization is weak but the ground state of the corresponding uniform chain is a spontaneously 
formed valence-bond-solid (where the spinons are deconfined). Based on our numerical results, we argue that a system with $\lambda \ll \Lambda$ is associated
with weak repulsive short-range spinon-spinon interactions. In principle, both the length scales $\lambda$ and $\Lambda$ can still be individually tuned from small to 
infinite (with $\lambda \le \Lambda$) by varying model parameters. In contrast, in the ladder system the two lengths are always similar, and this is the case also in the 
weakly dimerized systems when the corresponding uniform chain is in the critical phase. In these systems, the effective spinon-spinon interactions are purely attractive 
and there is only a single large length scale close to criticality, which is reflected in the standard spin correlations as well as in the spinon characteristics.
\end{abstract}

\date{\today}
 

\maketitle

\section{INTRODUCTION}

In one-dimensional (1D) strongly correlated systems, the emergence of fractional quantum numbers is a generic consequence of collective behaviors \cite{fracreview}.
In the exactly solvable critical $S=1/2$ antiferromagnetic (AFM) spin chain, the fundamental excitations are soliton-like quasiparticles (kinks and anti-kinks), 
called {\it spinons}, which carry spin $1/2$ \cite{Faddeev81,Haldane91}. Similar objects exist also in the valence-bond-solid (VBS) state stabilized by 
frustrated interactions \cite{Shastry81}. A bound state of spinons can be induced in the Heisenberg chain by an external magnetic field \cite{Muller81}.  
In higher dimensions, in systems with long-range AFM order, the fundamental excitations are magnons with spin $1$, as explained successfully by spin-wave 
theory \cite{Van58}. Spinon excitations are associated with spin-liquid ground states, which have no broken symmetries described by conventional local order 
parameters (but do have non-local, topological order) \cite{Lee06}. In two-dimensional (2D) AFM systems, deconfined spinons should emerge when a transition into a VBS state 
is approached, according to the theory of ``deconfined'' quantum-critical points \cite{Senthil04,Senthil05,Sachdev08}. 

The search for spinons has been a quest in experimental and theoretical condensed matter physics for decades, primarily because the fractionalization of excitations 
is a characteristic of exotic collective quantum many-body states, such as the spin liquids \cite{Lee06,Sachdev08,Lee08}. Moreover, in some cases the mechanism of 
confinement of spinons is a condensed-matter analog of the confinement of quarks in quantum chromodynamics. In this paper, building on a previous brief 
presentation \cite{Tang11a}, we will explore systems where confinement and deconfinement of spinons can be detected and characterized using large-scale quantum 
Monte Carlo (QMC) simulations in the valence-bond (VB) basis. We here focus on a range of different 1D systems but note that the same ideas have also already been 
applied to 2D systems in the context of deconfined quantum-criticality \cite{Tang13}. 

The starting point of our studies is the $S=1/2$ AFM Heisenberg chain, defined by the Hamiltonian
\begin{equation}
\label{eq:Heisenberg}
H = J \sum_{i=1}^N{\bf S}_i \cdot{\bf S}_{i+1},
\end{equation}
where the nearest-neighbor coupling $J>0$, $N$ is the total number of spins, and we apply periodic boundary conditions. We will add other interactions to this model
later, in order to bring the system to the different types of ground states mentioned above.

The ground state of the plain Heisenberg model (\ref{eq:Heisenberg}) can in principle be solved exactly by the Bethe-Ansatz approach \cite{Bethe31}, but in 
practice many of its salient features, such as the power-law decaying spin-spin correlations, were found using the bosonization method \cite{Luther75}. Reflecting the 
deconfined spinons, the lowest excited states of the Heisenberg model form bands of degenerate singlets and triplets \cite{Cloizeaux62,Yamada69,Muller81} with the
energy $\epsilon_1({q})$ as a function of the total momentum $q$ of the state being $\epsilon_1({q}) = (\pi/{2})J|\sin({q})|$, which was first calculated by des 
Cloiseaux and Pearson using the Bethe ansatz.\cite{Cloizeaux62}  Forming all possible combinations of two spinons propagating independently with fixed momenta, 
$\tilde q_1$ and $\tilde q_2$ with $q=\tilde q_1 + \tilde q_2$ gives a continuum above the lower bound and an upper bound given by $\epsilon_2(q) = \pi J|\sin({q}/{2})|$. 
A large spectral weight between these bounds (concentrated close to the lower bound because of matrix elements~\cite{Caux05}), which is detectable in inelastic 
neutron scattering  experiments \cite{Lake05}, is considered a good indicator of spinons in one dimension. 

The continuum spectrum of spinons has been observed in weakly coupled-chain 
compounds such as copper pyrazine dinitrate and KCuF$_3$ at zero magnetic field \cite{Stone03, Lake05}, while in none-zero magnetic fields incommensurate modes 
have been observed \cite{Stone03,Dender97}. In another chain compound, CuCl$\cdot2$(dimethylsulfoxide), there is an effective internal staggered magnetic 
field present, and spinon bound states have been observed \cite{Kenzelmann04}. In addition, in the spin ladder system (C$_5$H$_{12}$N)$_2$CuBr$_4$, it was
reported that the magnon could be fractionalized into spinons by tuning the external magnetic field \cite{Thielemann09}. The above experimental results can
be modeled using the Heisenberg Hamiltonian (\ref{eq:Heisenberg}) including the other effects mentioned above (external fields, inter-chain couplings). In addition
to neutron scattering, other experimental signals of spinons have also been proposed \cite{Zhou11}. So far, however, all the experimental probes give 
indirect information on the existence of spinons, and not much information on the properties of spinons other than their dispersion and excitation continuum.

Motivated by the on-going interest in the quantum physics of fractionalization, in this paper we are interested in exploring other aspects of spinons and their
confinement-deconfinement transitions. Using the QMC approach introduced in Refs.~\onlinecite{Tang11a,Banerjee10} and 
used in Ref.~\onlinecite{Tang13} to study 2D systems, we here explore 
a wider range of 1D systems where confinement and deconfinement can be studied systematically under various conditions. The method operates in a basis of VBs
(two-spin singlets) and unpaired spins and allows us to compute quantities defining the size of an isolated spinon as well as the size of an $S=1$ bound state. 
We also show that the same length scales appear in standard spin correlation functions, but are harder to access there in practice because the signal only appears 
in the differences between correlations in different spin sectors (and is therefore very noisy in QMC calculations of large systems).

The structure of the rest of the paper is as follows: In Sec.~\ref{sec:method}, we introduce the projector QMC method and calculate observables used to characterize 
spinons. in Sec.~\ref{sec:1D}, we present results for the $J$-$Q$ chain model \cite{Tang11a,Banerjee10},
which undergoes a quantum phase transition from the Heisenberg critical phase to a 
spontaneously symmetry-broken valence-bond solid (VBS). This system has deconfined spinon excitation in the entire range of the ratio $Q/J$ of the Heisenberg exchange 
$J$ and a multi-spin coupling $Q$. To achieve confinement, in Sec.~\ref{sec:j1j2q3} we introduce a staggered pattern of $J$-interactions, as recently done also in an 
investigation of spinons binding to a static impurity \cite{doretto09}. In Sec.~\ref{sec:ladder} we study spinon confinement when two Heisenberg chains are coupled 
to form a ladder. In Sec.~\ref{sec:crr}, we discuss the fact that the same length scales that appear in our VB-based definition of spinons can also be identified 
in the fine-structure of the spin-spin correlations in the higher-spin states, thus confirming that these length scales are not basis dependent and can be 
investigated using other methods as well. We summarize our work and discuss future prospects in Sec.~\ref{sec:conclusion}. 

\section{METHODS AND CALCULATED OBSERVABLES}
\label{sec:method}

We use VB projector QMC (VBPQMC) algorithm, which has been described in detail in Refs.~\onlinecite{Sandvik10,Wang10,Tang11a}. Here we first briefly
review the essential ideas underlying simulations of spin systems with this algorithm, and then focus on the definitions of spinon quantities and how to evaluate them. 

\subsection{VB basis and projector QMC method}
\label{sec:vbqmc0}

Searching for the ground state of a Hamiltonian $H$, we start with a ``trial''  wave function and write it as the linear superposition of all 
eigenstates of $H$ as
\begin{equation}
\label{eq:trial}
|\Psi\rangle_t = \sum_n c_n|\Psi_n \rangle.
\end{equation}
We then operate with $H$ a number $m$ times on this trial state to project out the ground state $|\Psi_0\rangle$;
\begin{equation}
\label{eq:pmc}
(-H)^m|\Psi\rangle_t = c_0 (-E_0)^m \left [|\Psi_0 \rangle + \sum_{n>0} \frac{ c_n}{c_0} \left (\frac{E_n}{E_0} \right )^m |\Psi_n \rangle \right],
\end{equation}
where, since normally $E_0<0$, we have added a minus sign in front of $H$. Provided that $|E_n/E_0| <1$ for all $n>0$, which can always be accomplished by
adding some negative constant to $H$, the ground state is projected out when $m \to \infty$.

\begin{figure}
\centerline{\includegraphics[angle=0,width=6.5cm, clip]{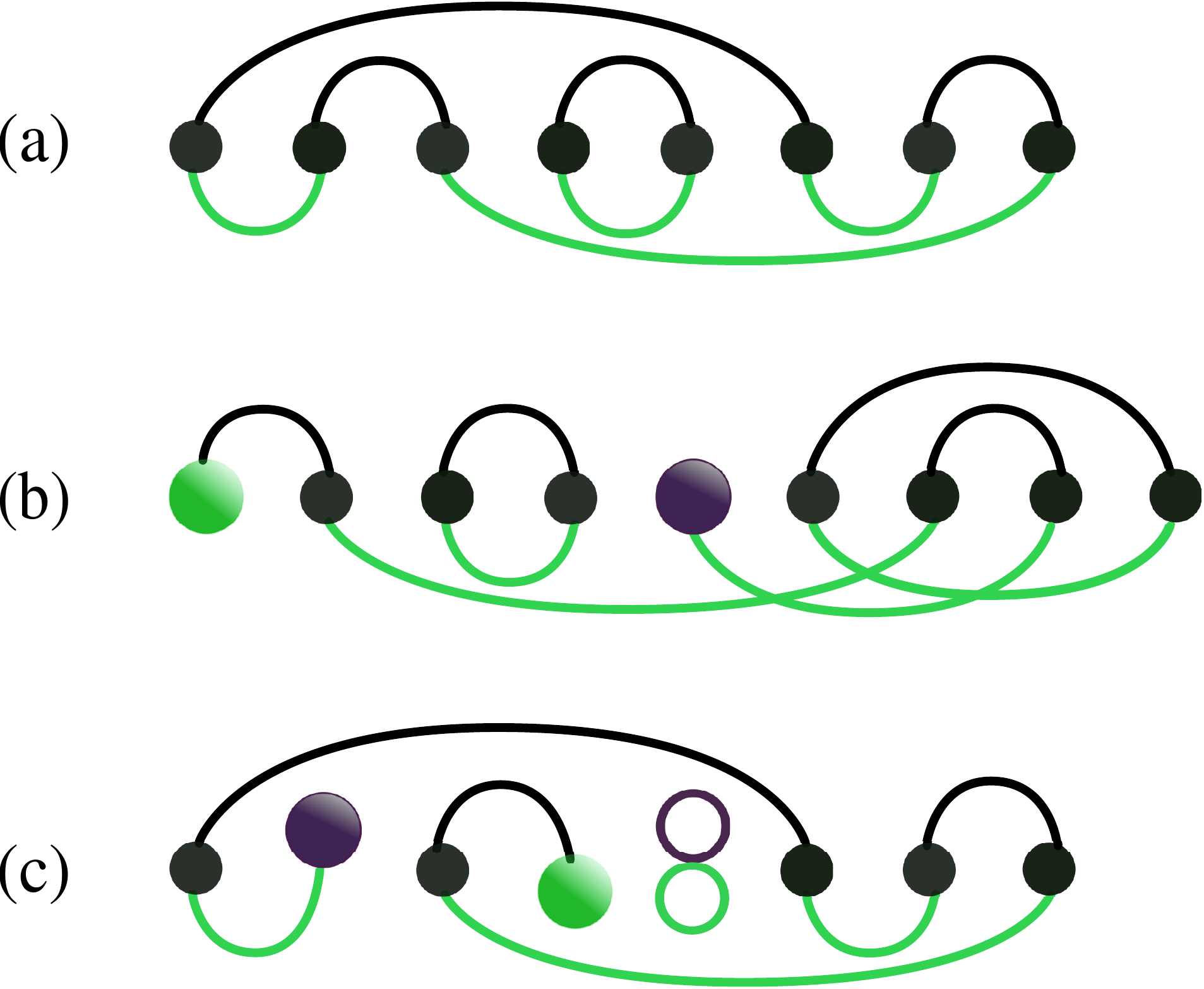}}
\caption{(Color online) Transition graph formed by bra (upper, black) and ket (lower, green) valence bond states on a spin chain. Part (a) shows an $S=0$
state on an even number of sites. In (b) the number of sites is odd and there is an unpaired spin in both the bra and the ket states. Part (c) shows an $S=1$
configuration, where there are two unpaired spins. In VBPQMC simulations, the distance distribution of the unpaired spins in (b) gives information on the 
size of an individual spinon, while the size of an $S=1$ bound state of two spinons is reflected in the distance distribution of unpaired spins on different 
sublattices in (c).}
\label{fig:loop}
\end{figure}

While the ground-state projection approach formulated above is completely general, 
the use of the VB basis has distinct advantages \cite{Sandvik05,Beach06}, as the spin of the trial state can be
chosen to match that of the ground state under investigation. For the bipartite spin models we are interested in here, if the number of spins $N$ is even, 
then the ground state is a singlet and a VB basis state can be written as
\begin{equation}
\label{vbstates}
|V_{\alpha}\rangle =  \prod_{i=1}^{N/2} |a,b\rangle_i, 
\end{equation} 
where $|a,b\rangle_i$ is the $i$th VB (singlet),
\begin{equation}
\label{singlet}
|a,b\rangle_i = \frac{1}{\sqrt{2}}\bigl (|\up_{a(i)} \dn_{b(i)}\rangle - |\dn_{a(i)} \up_{b(i)}\rangle \bigr), 
\end{equation} 
with $a(i)$ and $b(i)$ sites on sublattices $A$ and $B$, respectively. The trial state can be expanded in these VB basis states as
\begin{equation}
|\Psi\rangle_t = \sum_{\alpha} f_{\alpha} |V_{\alpha}\rangle,
\label{psitfa}
\end{equation}
where the coefficients $f_{\alpha} \ge 0$, reflecting Marshall's sign rule for the ground state of a bipartite system \cite{Liang88,Sutherland88}. 
It should be noted that the VB basis is 
overcomplete and, therefore, the expansion coefficients $f_\alpha$ are in principle not unique, which, however, is not explicitly of importance in the 
work discussed here. What is important is that the basis is non-orthogonal, with the overlap between two states given by \cite{Liang88,Sutherland88}
\begin{equation}
\label{overlap}
\langle V_\alpha | V_\beta \rangle \propto 2^{n_{\rm loop}-N/2},
\end{equation}
where $n_{\rm loop}$ is the number of loops in the {\it transition graph} formed when superimposing the bond configurations of $|V_{\alpha}\rangle$ and 
$|V_{\beta}\rangle$. An example with $n_{\rm loop}=2$ is shown in Fig.~\ref{fig:loop}(a). Expectation values of interest can normally also be expressed using
transition graphs, e.g., for studying the spin-spin correlation operator 
\begin{equation}
\hat{C}(r)= \frac{1}{N}\sum_{i=1}^N{\hat{\bf S}_i} \cdot {\hat{\bf S}_{i+r}},
\label{ssop}
\end{equation}
we need matrix elements of the form,
\begin{equation}
\label{eq:crr}
\frac{\langle V_\alpha |{\hat{\bf S}_i} \cdot {\hat{\bf S}_j}| V_\beta \rangle}{\langle V_\alpha | V_\beta \rangle}
= \left \{ \begin{array}{rl}  \pm 3/4, & i,j~{\rm in~same~loop,} \\ 0, & i,j~{\rm in~different~loops.} \end{array} \right.
\end{equation}
where the $+$ and $-$ sign in front of $3/4$ applies for sites on the same and different sublattices, respectively. 
Other examples of transition-graph estimators, e.g., dimer-dimer correlations of the form
\begin{equation}
\hat{D}_{xx}(r)= \frac{1}{N}\sum_{i=1}^N{(\hat{\bf S}_i} \cdot {\hat{\bf S}_{i+\hat{x}})(\hat{\bf S}_{i+r}} \cdot {\hat{\bf S}_{i+r+\hat{x}})},
\label{eq:drr}
\end{equation}
 \noindent have been discussed in Refs.~\onlinecite{Beach06} and \onlinecite{Tang11b}.

In the ``double projection'' version of the VBPQMC method \cite{Sandvik05}  that we use here, bra and ket VB states are generated stochastically by operating on 
the bra and ket versions of the trial state with strings of $m$ Hamiltonian terms (operators defined on bonds or groups of bonds for $J$ and $Q$ interactions, 
respectively). The probability of the bra $\langle V_\alpha|$ and ket $|V_\beta\rangle$ appearing together is given by
\begin{equation}
P_{\alpha,\beta}=g_\alpha g_\beta \langle V_\alpha|V_\beta\rangle,
\label{pab}
\end{equation}
where the unknown coefficients are such that
$\sum_\alpha g_\alpha |V_\alpha\rangle$ approaches the ground state of $H$ when $m\to \infty$ and expectation values in this ground state are obtained 
using the stochastically generated transition graphs $\langle V_\alpha|V_\beta\rangle$. For details of the computational procedures, which make use of very 
efficient loop updates, we refer to Ref.~\onlinecite{Sandvik10}.

For the trial state, we normally choose an amplitude-product state \cite{Liang88}, where the coefficients $f_\alpha$ in (\ref{psitfa}) are simple products of 
amplitudes $h_\alpha(r)$ corresponding to bond-lengths $r$;
\begin{equation}
f_\alpha = \prod_{i=1}^{N/2} h^{n_\alpha(r)}_\alpha(r),
\label{apcoeff}
\end{equation}
where $n_\alpha(r)$ is the number of bonds of length $r$ in VB configuration $\alpha$. These amplitudes can in principle be determined 
variationally \cite{Liang88,Lou07,Sandvik10} to optimize the trial state, but in practice such optimization is not crucial and the simulations converge well 
regardless of the details of the trial states. We typically choose a power-law form, e.g., $h_\alpha(r) = r^{-2}$. The bonds configurations of the trial 
state are sampled stochastically as well \cite{Sandvik10}.

Our VBPQMC calculation projects out the lowest state with given total spin, $S=0$ as discussed above or higher spins, as will be discussed further in the following. 
With periodic systems, the momentum is also a good quantum number and is determined by the trial state. With the simple amplitude-product trial states we are 
using, the momentum can be obtained very easily by translating the bonds by one lattice spacing. If the number of bonds is odd, i.e., the number of sites 
is of the form $N=4n+2$ for some integer $n$, this results in a negative phase, and, thus, the momentum $k=\pi$. Otherwise, for $N=4n$, there is
no phase and $k=0$. These are exactly the momenta of the ground states of bipartite spin chains.

\subsection{Generalized VB basis for $S>0$}
\label{sec:vbqmc1}

In addition to the use of the VB basis for singlet ground states, extensions of the VB basis with unpaired spins also provide a natural and convenient way to describe 
excitations with higher spin \cite{Wang10,Tang11a,Banerjee10}. In our study of spinons, we will study systems with one or two unpaired spins. In the former case, the total
number of sites $N$ is odd, and a generalized VB state can be written as
\begin{equation}
\label{eq:shalf}
|V_{\alpha}(r)\rangle =  \Big[ \bigotimes_{i=1}^{(N-1)/2} |a,b \rangle^{\alpha}_i \Big] \otimes |\up_r\rangle, 
\end{equation}
where the notation explicitly indicates the location $r$ in the chain of the unpaired spin and $\alpha$ labels the possible $(N-1)/2$-bond configurations
with this site excluded. For system with even $N$ and two unpaired spins, analogously an extended VB basis state is written as
\begin{equation}
\label{eq:s1}
|V_{\alpha}(r_a,r_b)\rangle =  \Big[ \bigotimes_{i=1}^{N/2-1} |a,b \rangle^\alpha_i \Big] \otimes |\up_{r_a}\rangle \otimes |\up_{r_b}\rangle, 
\end{equation}
with $N/2-1$ singlet pairs and two unpaired spins on different sublattices. These extended VB bases are also overcomplete and non-orthogonal
in their respective total-spin sectors $S$, and, if we choose (as we do here) the unpaired spins to have $S^z_i=1/2$, the $z$-projection of the
total spin is $S^z=S$.

The transition graphs shown in Figs.~\ref{fig:loop}(b) and \ref{fig:loop}(c) have open strings [with an open string of length zero being a special case corresponding to a
bra and ket spinon residing on the same site, an example of which is seen in Fig.\ref{fig:loop} (c)] in addition to loops. If we fix the spin-$z$ orientation of the unpaired 
spins, as we do here, the strings do not contribute to the weight (since they only have one allowed state, in contrast to the two allowed states of each loop)
and the overlap of two states is still given by Eq.~(\ref{overlap}). Note, in particular, that the unpaired spins can be at different lattice locations and 
the states still always have non-zero overlap. The strings do contribute to expectation values.

It should be pointed out that, in periodic chains of odd size $N$, which we use here to study a single unpaired spin in $S=1/2$ states, there is 
magnetic frustration caused by the boundary condition and the lattice is no longer strictly bipartite. Thus, maintaining the updating rules in the 
simulations \cite{Sandvik10,Sanyal11} the VB singlets here can some times be formed between sites on the same sublattices if we continue to label the sites as 
alternating A and B, except for one instance of adjacent AA or BB sites. (in the simulation we do not explicitly label the sites and there is no breaking of 
translational symmetry as we just use the same updating rules for the bonds and unpaired spins as for the even-$N$ chains). The distance between the unpaired 
spin in the bra and ket can then be an odd number of lattice spacings (while it is always even in a true bipartite chain). In many cases (which we will discuss 
in detail in Sec.~\ref{sec:1D}) the system is completely dominated by short bonds and the distance between the bra and  ket spinon is then always even in practice.

The trial states used for $S>0$ calculations are simple generalizations of the amplitude-product states discussed in Sec.~\ref{sec:vbqmc0}, with the 
wave-function coefficient given by Eq.~(\ref{apcoeff}) with no dependence on the unpaired spins. In principle one could improve the trial states by factors
depending on the unpaired spins and spin-bond correlations as well (as recently investigated in detail in Ref.~\onlinecite{Lin12}), but this is not necessary 
here. Following the reasoning in Sec.~\ref{sec:vbqmc0}, for $S=1$, $k=\pi$ for $N=4n$ and $k=0$ for $N=4n+2$, i.e., the momentum difference with respect 
to the $S=0$ ground state is $\pi$ in both cases, as it should be for the lowest triplet excitation. For the $S=1/2$ states, if we strictly label the
sites with sublattice labels $A$ and $B$, there is a defect in the odd-$N$ system, as discussed above. However, in the simulations there are no
explicit references to sublattices and in effect the system is then translationally invariant. Then, under the further assumption that no bonds with 
length as large as $N/4$ are present (such configurations having ill-defined signs) \cite{Shao15}, the momentum is $k=0$ or $\pi$, for $N$ of the 
forms $4n+1$ and $4n+3$, respectively.

\subsection{Characterization of spinons in the VB basis}

In order to study spinon sizes and confinement lengths, we consider overlaps written in the form
\begin{equation}
\label{eq:overlapShalf}
_{\frac{1}{2}}\langle \Psi_0 | \Psi_0 \rangle_{\frac{1}{2}} = \sum_{r,r'} \sum_{\alpha, \beta} g_{\alpha}(r)g_{\beta}(r') \langle V_{\alpha}(r)| V_{\beta}(r') \rangle, 
\end{equation}
generalizing Eq.~(\ref{pab}) to $S=1/2$ (single-spinon) systems and written explicitly using sums of terms with all possible locations of the unpaired spins. We
have an analogous form
\begin{align}
\label{eq:overlapS1}
\begin{split}
_{1}\langle \Psi_0 | \Psi_0 \rangle_{1} =  &\sum_{r_a,r_b}\sum_{r'_a,r'_b} \sum_{\alpha, \beta} g_{\alpha}(r_a,r_b)g_{\beta}(r'_a,r'_b) \\
&\langle V_{\alpha}(r_a, r_b)| V_{\beta}(r'_a, r'_b) \rangle,
\end{split}
\end{align}
for $S=1$ (spinon-pair) systems. 

The overlaps are not computed explicitly in the simulations but serve as normalization factors and weights in the sampling procedures, such that 
the different contributions to the above sums appear according to their relative weights. The practical simulation procedures for $S>0$ are relatively 
straight-forward generalizations of the method with loop updates for $S=0$. We refer to Refs.~\onlinecite{Banerjee10,Wang10,Tang11b} for technical details. 
In the following, we discuss distribution functions used to characterize spinons. We will here make us of the unpaired spins, although in principle one can also
define spinon quantities using the entire strings, of which the unpaired spins are the end points.

\subsubsection{Single-spinon distribution function}

As discussed above, in the VBPQMC method the bra and ket states are generated stochastically, and for $S=1/2$ we can use Eq.~(\ref{eq:overlapShalf})
to define a distribution of the separation of the unpaired spins in the bra and ket states. Restricting ourselves to a translationally invariant system we 
have the probability of separation $r-r'$ (up to an irrelevant normalization factor which is easily computed at the end):
\begin{equation}
P_{AA}(r-r') = \sum_{\alpha, \beta} g_{\alpha}(r)g_{\beta}({r'}) \langle V_{\alpha}(r)| V_{\beta}(r') \rangle,
\label{paadef1}
\end{equation}
where the subscript $AA$ serves to indicate that the unpaired spins should be on the same sublattice (because there is an excess of one site on one of the
sublattices, which is the sublattice with the unpaired spin), which we can take as the $A$ sublattice. Thus,
$P_{AA}(r)$ should vanish when the separation $r$ is an odd number of lattice spacings. Our basic assertion is that, if spinons are well-defined quasiparticles 
of the system, then we expect $P_{AA}$ to reflect the size and shape of an {\it intrinsic} ``wave packet'' within which the net magnetization $S^z=1/2$ carried
by the spinon is concentrated. We will show in the following that 1D VBS states are characterized by an exponentially decaying overlap, $P_{AA}(r) \propto e^{-r/\lambda}$, 
and it is then natural to take $\lambda$ as a definition of the intrinsic spinon size.

We should here note again that, for a periodic system with an odd number of sites, there is, strictly speaking, no absolute distinction between the sublattices
(i.e., the system is strictly speaking not bipartite). However, when the system size $N \to \infty$ we in general expect the role of the boundary condition 
to diminish and $P_{AA}(r)$ to tend to zero for any given odd $r$. In Sec.~\ref{sec:1D}, we will discuss in detail how this limit is approached, and we will also 
see an example (one where spinons are not well-defined quasi-particles) where the boundaries continue to play a role even for infinite size.

\subsubsection{Two-spinon distance distribution function}

In the case of $S=1$ states (two spinons), we can define several different distributions. Here, we will focus on the separation of spinons on {\it different} 
sublattices in the bra and ket;
\begin{eqnarray}
&& P_{AB}(r_a-r'_b) = \sum_{\alpha, \beta}\sum_{r_b,r'_a}  g_{\alpha}(r_a,r_b)g_{\beta}(r'_a,r'_b) \times \nonumber \\
&& ~~~~~~~\langle V_{\alpha}(r_a,r_b)| V_{\beta}(r'_a,r'_b) \rangle.
\label{pabdef}
\end{eqnarray}
In the case where a single spinon is a well-defined quasi-particle, i.e., $\lambda < \infty$, we expect this quantity to give us information on the
confinement or deconfinement of two spinons. In the former case, we will see that asymptotically $P_{AB}(r) \propto e^{-r/\Lambda}$ and, thus, we consider $\Lambda$ as a
definition of the confinement length-scale (i.e., the size of the $S=1$ spinon bound state). We will see that deconfined spinons give rise to characteristic
broad distributions.

We could also have defined the above distance distribution with the two unpaired
spins both in the bra or in the ket, and we have also investigated it. This distribution typically does not differ significantly from the one 
defined in Eq.~(\ref{pabdef}).

\subsubsection{Same-sublattice distribution in two-spinon states}

We will also study the analog of the $S=1/2$ quantity $P_{AA}(r)$ [Eq.~(\ref{paadef1})] in the triplet state, defined as
\begin{eqnarray}
&&P^*_{AA}(r_a-r'_a) = \sum_{\alpha, \beta}\sum_{r_b,r'_b}  g_{\alpha}(r_a,r_b)g_\beta(r'_a,r'_b) \times \nonumber \\
&& ~~~~~~~\langle V_{\alpha}(r_a,r_b)| V_{\beta}(r'_a,r'_b) \rangle,
\label{paadef2}
\end{eqnarray}

where we use the $*$ superscript to distinguish this distribution from the single-spinon distribution (\ref{paadef1}). We can define $P^*_{BB}$ in the same way, 
and use $P^*_{AA}(r)=P^*_{BB}(r)$ to improve the statistics. We will see that, under certain conditions, $P^*_{AA}$ of the triplet state contains the same 
information for the spinon size $\lambda$ as the $S=1/2$ quantity $P_{AA}$, and we can use this property of the $S=1$ state to characterize the intrinsic 
spinon size also in cases where the $S=1/2$ state breaks translational invariance and is not appropriate for use with our calculations presuming
translational invariance (the $2$-leg ladder system being such an example, which will be studied below in Sec.~\ref{sec:ladder}).

\section{Deconfined spinons in uniform spin chains}
\label{sec:1D}

We here first test the concepts and methods for a class of spin chains, the $J$-$Q_3$ model, which can be tuned between a ground-state phase with
properties similar to the standard critical Heisenberg chain and a VBS phase with VBs crystallizing on alternating nearest-neighbor bonds. In the critical 
state, spinons are rigorously known to be elementary excitations based on the exact Bethe-ansatz wave function of the plain Heisenberg chain, and in a VBS 
state there are also strong arguments for spinons \cite{Shastry81}. In either case, a pair of spinons can be regarded as a kink and an anti-kink of an ordered 
(in the case of the VBS) or quasi-ordered (in the critical state) medium. There is no apparent confining potential between these defects in one dimension (and 
clearly any effectively attractive potential would lead to a bound state and confinement of the spinons in the ground state, although deconfinement could still
take place at higher energy). Our calculations show explicitly that there are instead weak {\it repulsive} interactions, the effects of which diminish with the system 
size, thus leading to independently propagating spinons in the thermodynamic limit down to the lowest energies. We will also investigate a modified $J$-$Q_3$ model
with long-range interactions, which hosts a N\'eel ordered ground state. Here, spinons are not expected to be deconfined and we investigate the break-down
of the spinon as well-defined quasi-particle in this case.

\subsection{Results for the $J$-$Q_3$ chain}
\label{sec:jq3}

We here consider the 1D $J$-$Q_3$ chain Hamiltonian \cite{Tang11a}, 
\begin{equation}
\label{eq:jq3}
H = -\sum_{i}^N (J C_{i,i+1} + Q_3 C_{i,i+1}C_{i+2,i+3}C_{i+4,i+5}),
\end{equation}
where $C_{ij}$ is a singlet-projection operator on two sites,
\begin{equation}
C_{i,j}=1/4-{\bf S_i}\cdot{\bf S_j},
\end{equation}
and the $J$ term is simply the standard antiferromagnetic Heisenberg interaction. We here use the $Q_3$ term with three projectors, as its ground state
at the extreme point $J=0$ is more strongly VBS-ordered than that of the $Q_2$ model with only two projectors.

When the coupling ratio $g=Q_3/J$ is small, the system remains in the Heisenberg-type critical state, where the spin-spin correlation function $C(r)$, i.e., 
the expectation value of Eq.~(\ref{eq:crr}), has the asymptotic form $C(r) \sim \ln^{1/2}(r)/r$ \cite{Luther75,Giamarchi89, Afflect89}.
When $g$ is large, the $Q_3$ term enforces VBS ordering and $C(r)$ is exponentially decaying. The VBS state is two-fold degenerate. The physics of this
phase transition is identical (in the sense of universality) \cite{Tang11a,Sanyal11} to that in the frustrated $J_1$-$J_2$ chain, where spinons in the 
VBS state were discussed on the basis of a variational state by Shastry and Sutherland \cite{Shastry81, Byrnes99}. In field-theory language, the phase transition is driven 
by the sign-change of a marginal operator, and this operator is also the root cause of the logarithmic correction to $C(r)$ in the critical phase. Exactly at the
critical--VBS transition point the correlations decay as $1/r$ with only very small corrections. The transition point of the $J$-$Q_3$ model is at
$g_c=(Q_3/J)_c \approx 0.1645$, as determined from level spectroscopy \cite{Tang11a} (excited-state singlet-triplet crossing \cite{Eggert}) and VBPQMC 
calculations of correlation functions \cite{Sanyal11}. 

Although we do not expect the Hamiltonian (\ref{eq:jq3}) to be naturally realizable in any specific material, the fact that it has the same kind of ground state phases as
the more realistic frustrated  $J_1$-$J_2$ chain still makes its physics interesting, and not being frustrated in the standard sense it is not associated with sign problems in QMC 
simulations. The same physics of spontaneous dimerization also occurs in spin chains with phonons (often called spin-Peierls systems) \cite{Suwa15}.
We expect the properties of spinons to be discussed below to apply also to frustrated chains and spin-Peierls systems.

\subsubsection{Single spinons in states with total-spin $1/2$}

\begin{figure}
\centerline{\includegraphics[angle=0,width=7.5cm, clip]{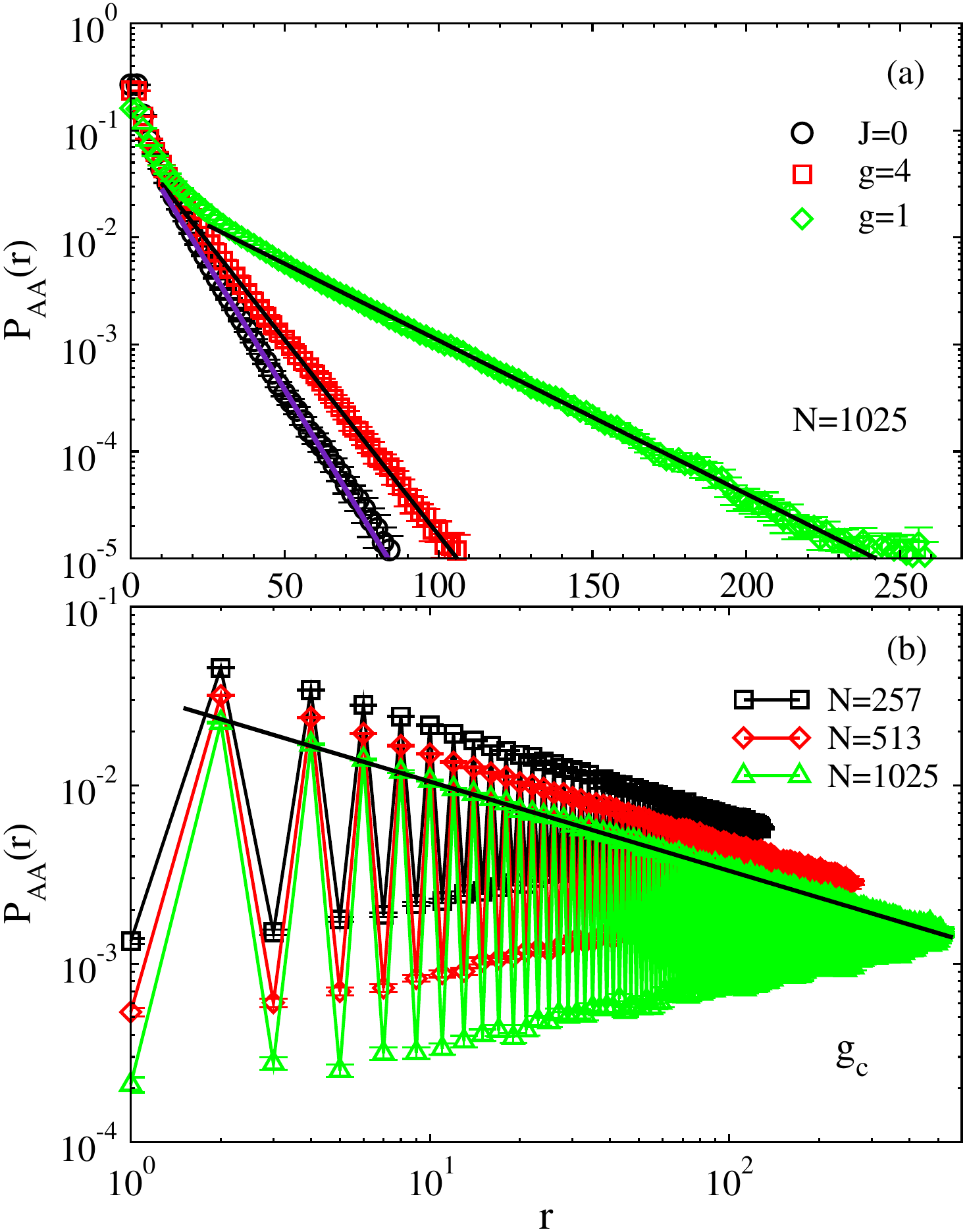}}
\caption{(Color online) Single spinon overlap distribution in the $J$-$Q_3$ chain. (a) Exponential decays indicating well-defined quasi-particles in VBS states at
different values of $g=Q_3/J$. The size $\lambda$  of the spinon (the inverse of the slopes of the lines on the lin-log plot) diverges as the critical point 
is approached. Panel (b) shows that the spinon is marginally defined at the critical point, with the overlap decaying as a power-law with exponent $\alpha=0.500(2)$
(with a fitted line to the even-$r$ points shown for $N=1025$). The even-odd oscillations are due to the frustration caused by the single-spinon defect in a periodic 
chain (with the odd-$r$ contributions only possible in a non-bipartite system). The effects of frustration for $r$ less than $N/2$ diminish as the chain size increases.}
\label{figa1}
\end{figure}

We here first investigate $P_{AA}(r)$ as defined in Eq.~(\ref{paadef1}) to study the size of spinons in the VBS phase at different coupling ratios $g=Q_3/J$. 
In Fig.~\ref{figa1}(a), we see that the intrinsic spinon wave packet has a pronounced exponential decaying form, $P_{AA}(r) \propto e^{-r/\lambda}$, showing that 
spinons indeed are well-defined quasi-particles of the VBS, with a characteristic size $\lambda$. The spinon size decreases with increasing $g$ (going deeper 
into the VBS phase), with $\lambda=30.0(1)$ when $g=1$ and $\lambda=9.2(1)$ when $g \rightarrow \infty$  (the pure $Q_3$ model). When $\lambda$ is large, there 
are also significant deviations from the pure exponential form for a range of small $r$, indicating cross-over behaviors to a different form obtaining
when $g \to g_c$. As shown in Fig.~\ref{figa1}(b), exactly at the transition point $g_c$ the decaying form is indeed no longer exponential, instead it is 
very well described by $r^{-\alpha}$ with the power $\alpha = 0.500(2)$. Our physical interpretation of this result is that, the spinon at the transition point 
can be considered only as a marginally well-defined quasi-particle in real space. 

\begin{figure}
\centerline{\includegraphics[angle=0,width=7.5cm, clip]{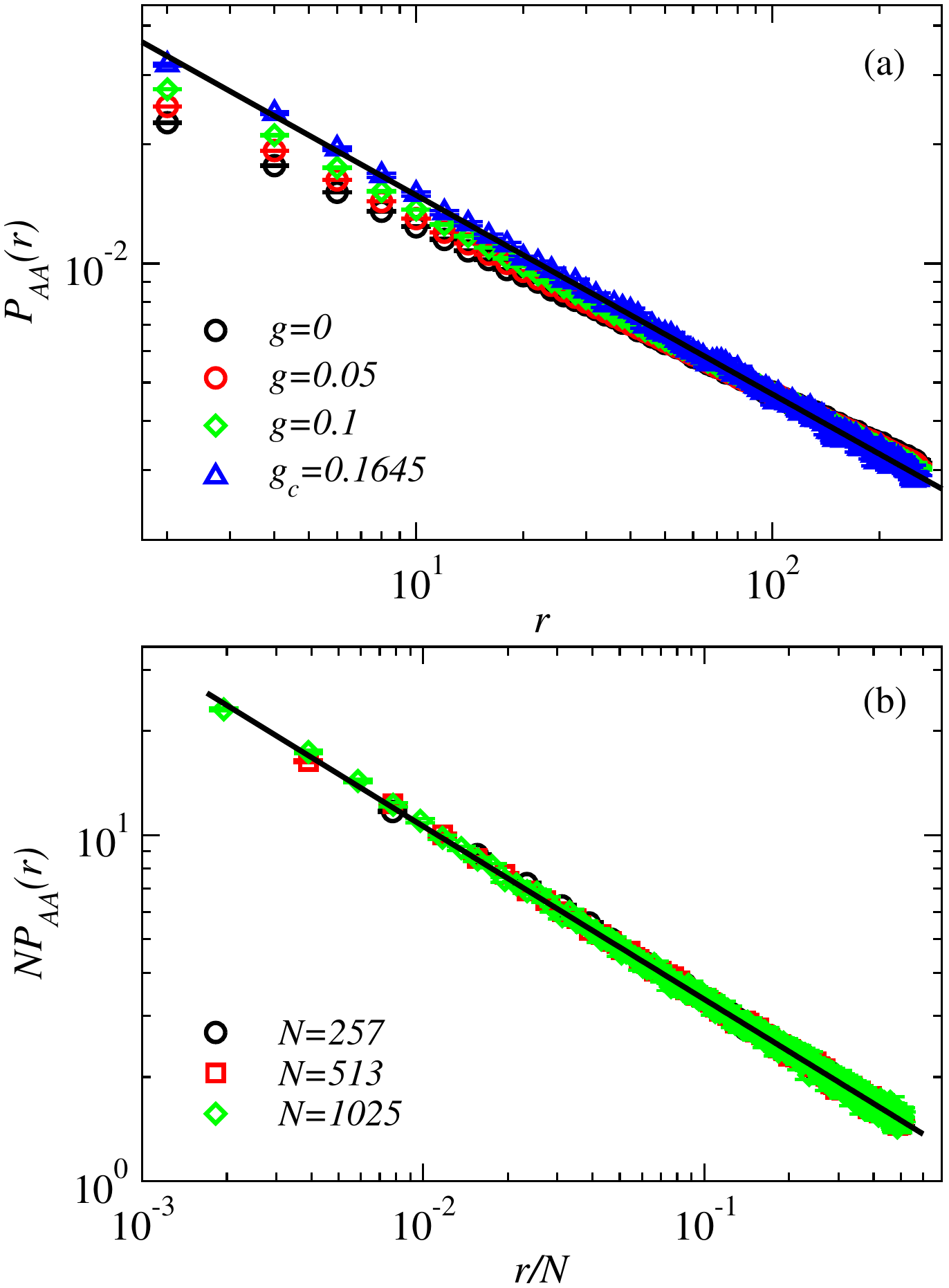}}
\caption{(Color online) (a) Single-spinon distribution function at the VBS transition point and inside the critical phase ($g \le g_c =0.1645$) computed using 
chains of length $N=513$. (b) The data at $g_c$ for several system sizes, rescaled such that data collapse is achieved. The lines in both (a) and (b) correspond 
to the $r^{-1/2}$ form.}
\label{figa5}
\end{figure}

As we discussed in Sec.~\ref{sec:vbqmc1}, for $N$ odd there is a complication with the periodic boundaries, which renders the system non-bipartite in principle. 
The distance between the unpaired spin in the bra and ket can then be odd. However, the probability of these odd distances is exceedingly small in the VBS state of 
the $N=1025$ chains used in Fig.~\ref{figa1}(a), but in the critical-chain results in Fig.~\ref{figa1}(b) we clearly can see non-zero odd-$r$ probabilities. 
Relative to the even-$r$ probabilities, for fixed $r$ they decrease rapidly as $N$ grows, while approaching the even-$r$ probabilities as $r \to N/2$ (and, 
interestingly, the odd branch follows almost an inverse of the behavior of the even branch, increasing as $r^{-0.5}$ in the relevant range of $r$). In our simulations 
we neglect the non-trivial (non-Marshall) 
signs in the wave function arising from the even-length bonds (where we define the length as the shortest of the two possible distances between the two paired
spins under the periodic boundary conditions), but we find it unlikely that this approximation would affect our conclusions on the nature of the spinon as 
these signs also are due to boundaries and we are interested in the thermodynamic limit. We will also see further in what follows that we obtain the same exponential
(for $g<g_c$) or power-law (for $g=g_c$) decay also in $P^*_{AA}$ [Eq.~(\ref{paadef2})], in the chains with two unpaired spins, where the lattice remains 
bipartite and there are no frustration effects.

Given the fact that the exponent $\alpha$ of the critical spinon overlap in Fig.~\ref{figa1}(b) is very close to $1/2$, and the behavior is seen to remarkable 
consistency over two orders of magnitude of $r$, we conjecture that the exponent should in fact be exactly $1/2$. It is tempting to associate it with the square-root 
of the spin correlation function $C(r)=1/r$, although we have not tried to formally compute this quantity within the bosonization approach (which in
principle should be possible \cite{Giamarchi}).

\begin{figure}
\centerline{\includegraphics[angle=0,width=7.5cm, clip]{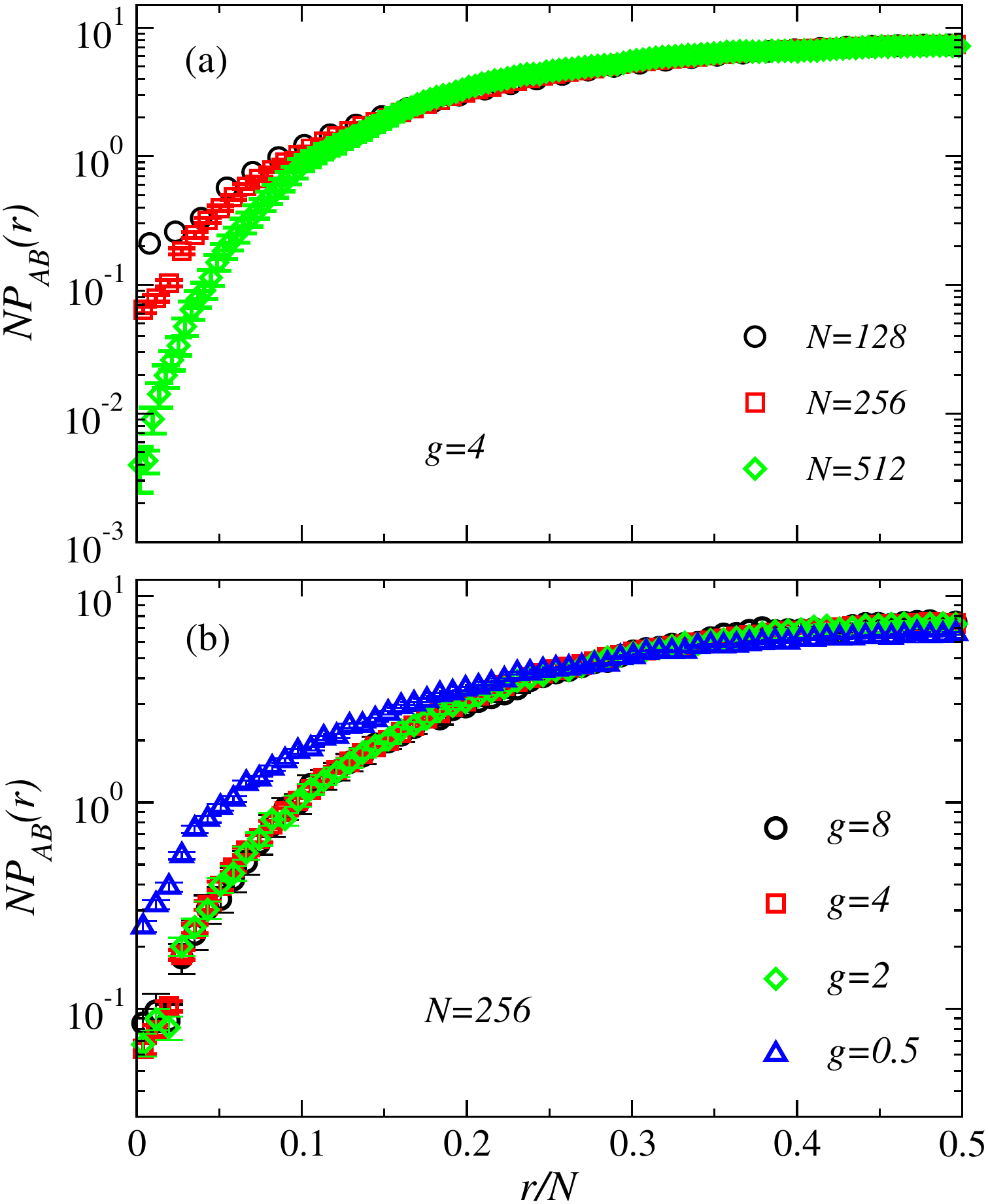}}
\caption{(Color online) Two-spinon distance distribution in VBS states of the $J$-$Q_3$ chain at (a) fixed $g=4$ and different chain lengths, and (b) fixed 
chain length $N=256$ and different coupling ratios. The $y$ and $x$ axes have been rescaled with $N$ and $1/N$, respectively, in order to achieve data collapse
for large $r$ in (a). The increase in the small-$r$ distribution for the lowest $g$-value in (b) shows that the effective short-distance spinon-spinon repulsion becomes 
weaker as the system approaches the the transition point ($g_c=0.1645$).}
\label{figa3}
\end{figure}

Another interesting question to ask is as follows: How is the critical $\sim r^{-1/2}$ form of the single-spinon distribution $P_{AA}(r)$ at $g_c$ changed when 
going further into the critical region ($g < g_c$)? The logarithmic correction to the correlation function $1/r$ is a well known consequence of the presence
of a marginal operator, as mentioned above. One would then expect corrections to $P_{AA}(r)$ as well. As seen in Fig.~\ref{figa5}(a), $P_{AA}(r)$
indeed changes noticeably when moving away from the transition point into the $g<g_c$ critical phase. The behavior can be fitted to a power-law with
exponent depending on $g$, but most likely the $r^{-1/2}$ behavior persists for all $0 \le g \le g_c$ and it is only the strength of a logarithmic
correction that changes. While the data can be fitted to the $r^{-1/2}$ with a multiplicative logarithmic correction, the power of the logarithm is not
clear, and further quantitative studies of this behavior would require much longer chains. 

In Fig.~\ref{figa5}(b), we further analyze the behavior at $g_c$ for different system sizes, re-graphing the even branch of Fig.~\ref{figa1}(b) such that 
data collapse is achieved: $NP_{AA}$ versus $r/N$. 
An interesting aspect of these results is that there are no noticeable enhancements due to the periodic boundaries at the longest 
distances, $r \sim N/2$ (which are typically seen prominently in correlation functions), with the power law describing the data very well from the smallest to 
largest distances for all system sizes.

\subsubsection{Two spinons in states with total spin $1$}

\begin{figure}
\centerline{\includegraphics[angle=0,width=7.5cm, clip]{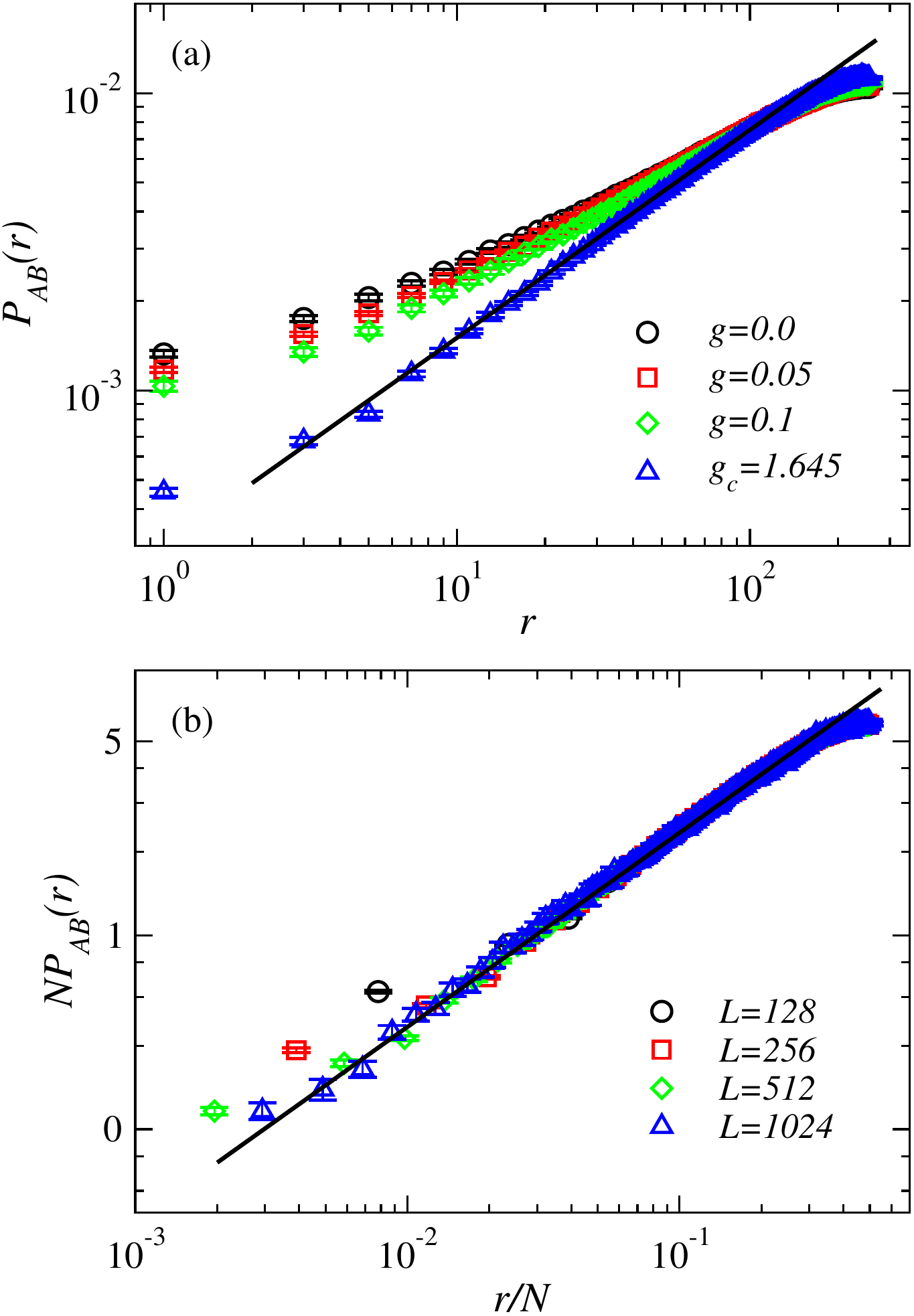}}
\caption{(Color online) Distribution of spinon separations in $S=1$ states at and below the VBS transition point $g_c$; in (a) for fixed
chain-length $N=512$ and varying $g$, and in (b) at $g_c$ for different chain lengths. The lines going through the $g_c$ points have slope $0.7$.}
\label{figabbelow}
\end{figure}

Next, we consider chains with even $N$ and two unpaired spins. The distribution function $P_{AB}(r)$ here reflects the effective mutual interaction between two 
spinons, mediated by the background of singlets. For a confining case, we would expect to observe $P_{AB}(r) \propto e^{-r/\Lambda}$, with a finite confinement length 
$\Lambda$. Deconfinement should be signaled by a divergence of $\Lambda$. Results for the $J$-$Q_3$ chain in the VBS phase, graphed in Fig.~\ref{figa3}, show 
distribution functions with no decay at long distances. Instead, $P_{AB}(r)$ exhibits a very broad maximum at the largest distance, which we naturally interpret as 
resulting from a weak repulsion between two spinons. As shown in Fig.~\ref{figa3}(a), the repulsion diminishes somewhat when tuning down the coupling ratio toward 
the critical point, where, apparently, increasing quantum fluctuations (including an increasing fraction of long VBs) reduce the repulsive potential. The range of 
$r$ over which the distribution is almost flat increases essentially proportionally with $N$. In Fig.~\ref{figa3}(a), we have multiplied the distribution function with 
$N$ for several $N$ at a fixed $g$ inside the VBS phase, and find that the curves collapse well on top of each other for $r/N$ roughly in the range $0.1$ to $0.5$. 
This indicates that the effective interactions are short-range in nature, with spinons far away from each other behaving as free particles. Clearly, all these results 
point to deconfined spinons, as expected. While the details of the cause of the repulsive potential are uncertain, it is clear that the sign of the effective 
interaction is crucial for deconfinement (at the lowest energies studied here); any weak attractive potential would bind the spinons, while short-range repulsive interactions aid deconfinement.

\begin{figure}
\centerline{\includegraphics[angle=0,width=8cm, clip]{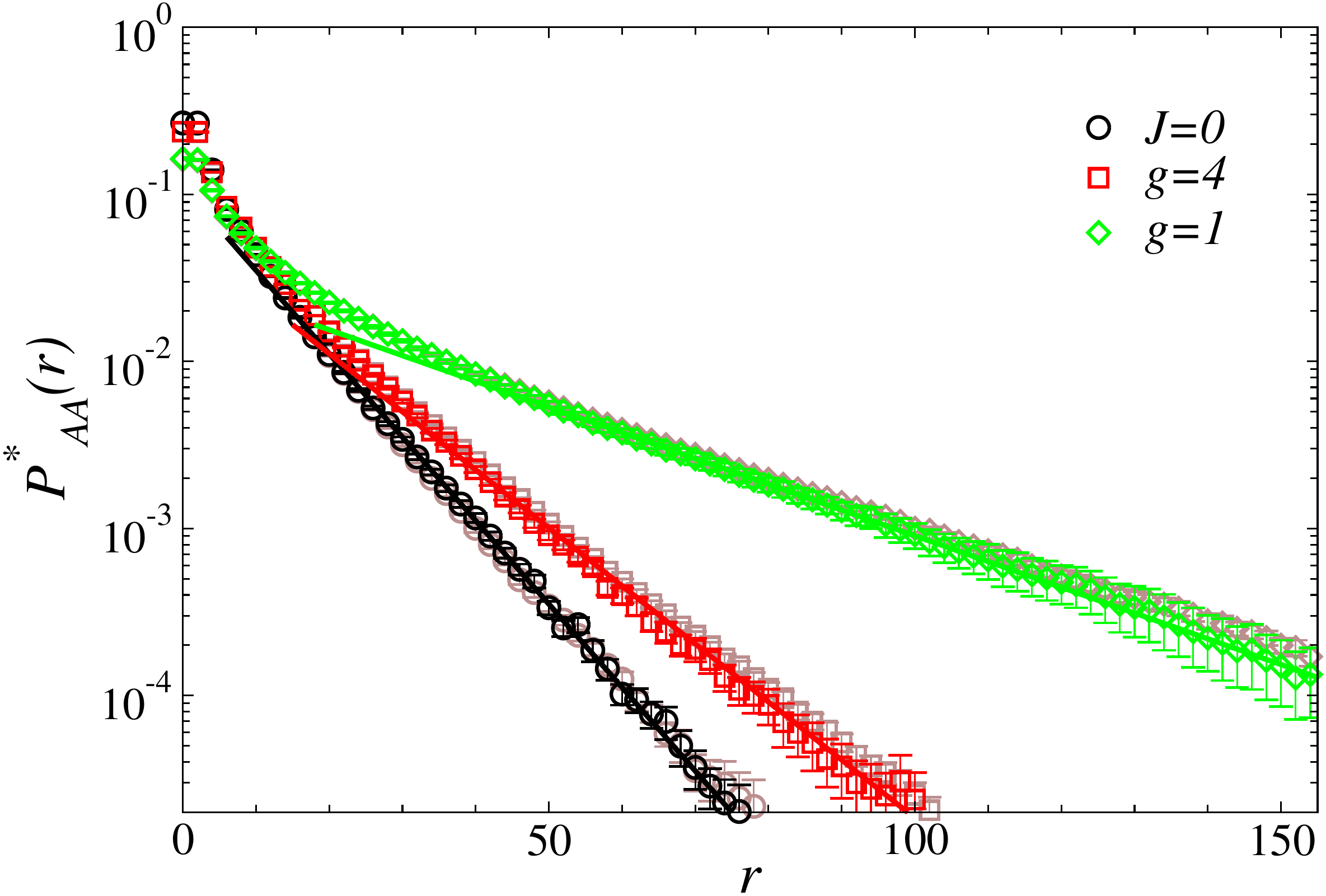}}
\caption{(Color online) The same-sublattice distribution function for $S=1$ states at three different values of the coupling ratio.
The corresponding distributions $P_{AA}(r)$ for the $S=1/2$ states at the same couplings are shown in lighter (brown) color and they
coincide very closely with the $S=1$ functions (thus, demonstrating that the single-spinon size can be obtained also from the $S=1$
simulations). The system size here is $N=1024$ for $S=1$ and $1025$ for $S=1/2$.}
\label{figpaaplus}
\end{figure}

Results for $P_{AB}(r)$ at the VBS transition and inside the critical phase are shown in Fig.~\ref{figabbelow}(a), while results for several chain
lengths at the critical point are shown with rescaled axis to achieve data collapse in \ref{figabbelow}(b). The critical distribution is also here consistent with a power-law, $P_{AB}(r) \sim r^\gamma$, with $\gamma \approx 0.7$ (and with a prefactor decreasing with the system size). Based on these results one may argue that the effective spinon-spinon interactions become increasingly long-ranged as $g_c$ is approached from the VBS side, although the short-range part is decreasing, based on the fact that distribution at short distances grows upon decreasing $g$. Inside the critical phase there are again likely logarithmic corrections, and the trend of decreasing effective short-distance spinon-spinon interactions continues as $g$ decreases.

Next, we consider the same-sublattice distribution function $P^*_{AA}(r)$, defined in Eq.~(\ref{paadef2}). Since the spinons are deconfined and typically are further
away from each other than the single-spinon length-scale $\lambda$, one would expect that $P^*_{AA}(r)$ contains essentially the same information as the single-spinon 
function $P_{AA}(r)$ for the $S=1/2$ state, defined in Eq.~(\ref{paadef1}). This is indeed the case in the VBS phase, as demonstrated in Fig.~\ref{figpaaplus}. 
Clear exponential decays are observed, and the results coincide almost perfectly with the previous results for $P_{AA}(r)$ in Fig.~\ref{figa1}(a). 

To reiterate what is going on here, the two spinons in the $S=1$ state are on different sublattices, and the unpaired spin on sublattice A in the ket state is 
correlated to the one on the same sublattice in the bra state, to within the length-scale $\lambda$ that we have argued describes the internal spinon size. 
The same holds for the unpaired bra and ket spins on sublattice $B$. Due to spinon deconfinement the $A$ and $B$ spinons are not bound to each other, however, 
and typically are far away from each other. Under these conditions, the distribution functions $P_{AA}(r)$ and $P^*_{AA}(r)$ are essentially the same.

\begin{figure}
\centerline{\includegraphics[angle=0,width=8cm, clip]{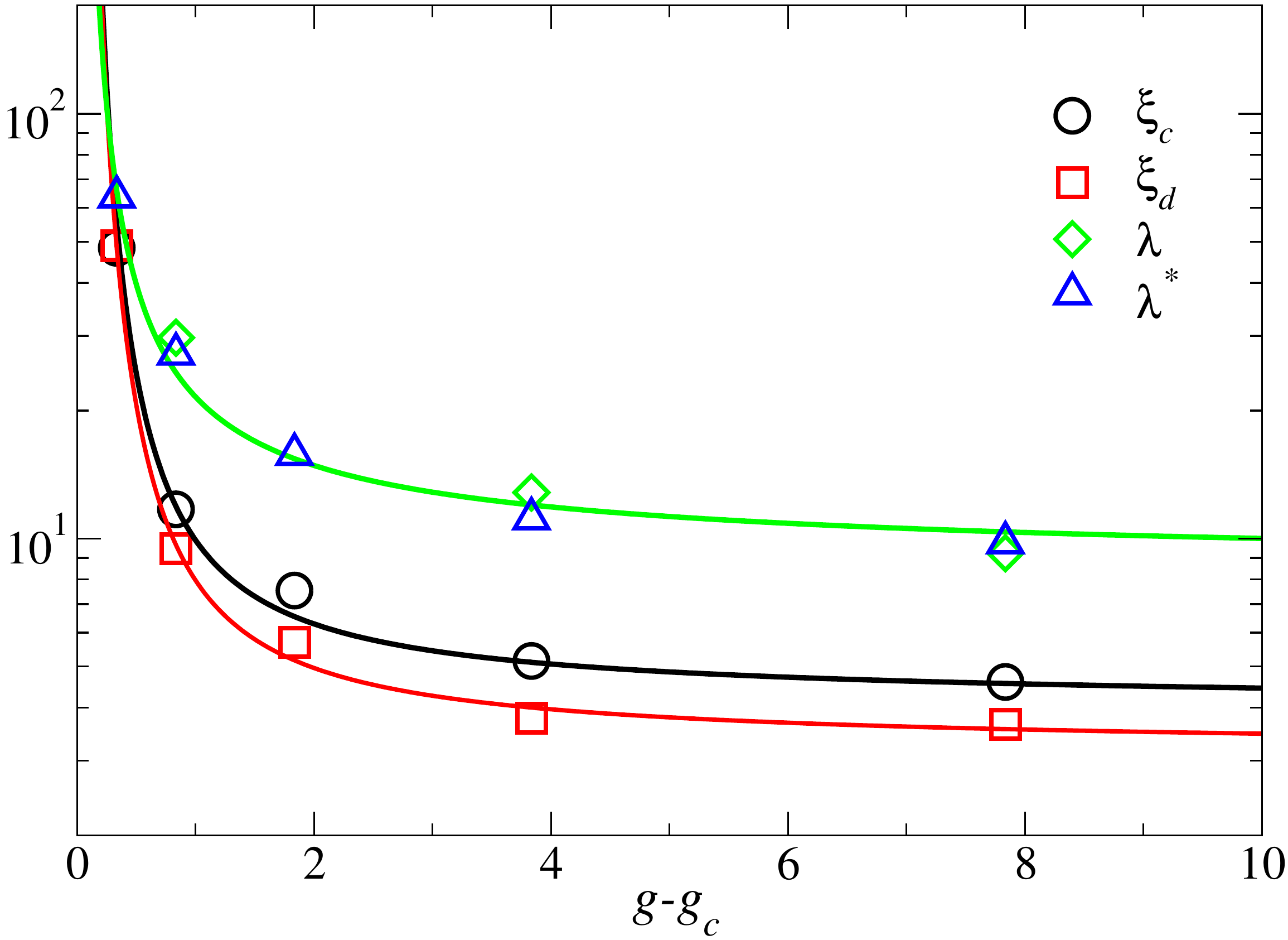}}
\caption{ (Color online) Spin and dimer correlation lengths, $\xi_c$ and $\xi_d$, along with the spinon size measured in the $S=1/2$ and $1$ states, $\lambda$ and $\lambda^*$, upon approaching the critical point $g_c=0.1645$ from the VBS phase in 1D $J$-$Q_3$ model. Since this transition is of the KT type, we fit the data to the form $ae^{b/\sqrt{g-g_c}}$ (solid lines).}
\label{exp3}
\end{figure}

To illustrate this point more explicitly, in Fig.~\ref{exp3} we plot results in the VBS state and approaching the critical point for the spinon-size estimates $\lambda$ and $\lambda^*$ [extracted from the distribution functions $P_{AA}(r)$ and $P^*_{AA}(r)$], together with the standard spin correlation length $\xi_c$ [obtained from the spin-spin correlation function (\ref{eq:crr})] and the VBS correlation length $\xi_d$ [extracted from dimer-dimer correlation function (\ref{eq:drr})]. It can be seen that $\lambda$ and $\lambda^*$ are almost identical to each other, as expected. The four lengths: $\xi_c$, $\xi_d$, $\lambda$, $\lambda^{*}$, diverge at a similar rate upon approaching the critical point $g_c=0.1645$. Since the phase transition from the ordered VBS state to the critical state in the 1D $J$-$Q_3$ model is similar to a 2D classical Kosterlitz-Thouless (KT) transition, we fit these four lengths with functions to the form of the correlation length in that case, $\xi \sim ae^{b/\sqrt{g-g_c}}$, where $a, b$ are fitting parameters. Due to the statistical errors and the small number of data points, we cannot determine these fitting parameters very precisely. Representative curves from these fits are shown in Fig.~\ref{exp3}. We also notice in Fig.~\ref{exp3} that the spinon size $\lambda$ extracted this way is much larger than the correlation lengths $\xi_c$ and $\xi_d$, which we will discuss again later in Sec.~\ref{sec:crr}, in connection with the correlation functions in $S=1/2$ or $1$ states (which, we argue, should also contain the spinon size).

\begin{figure}
\centerline{\includegraphics[angle=0,width=8cm, clip]{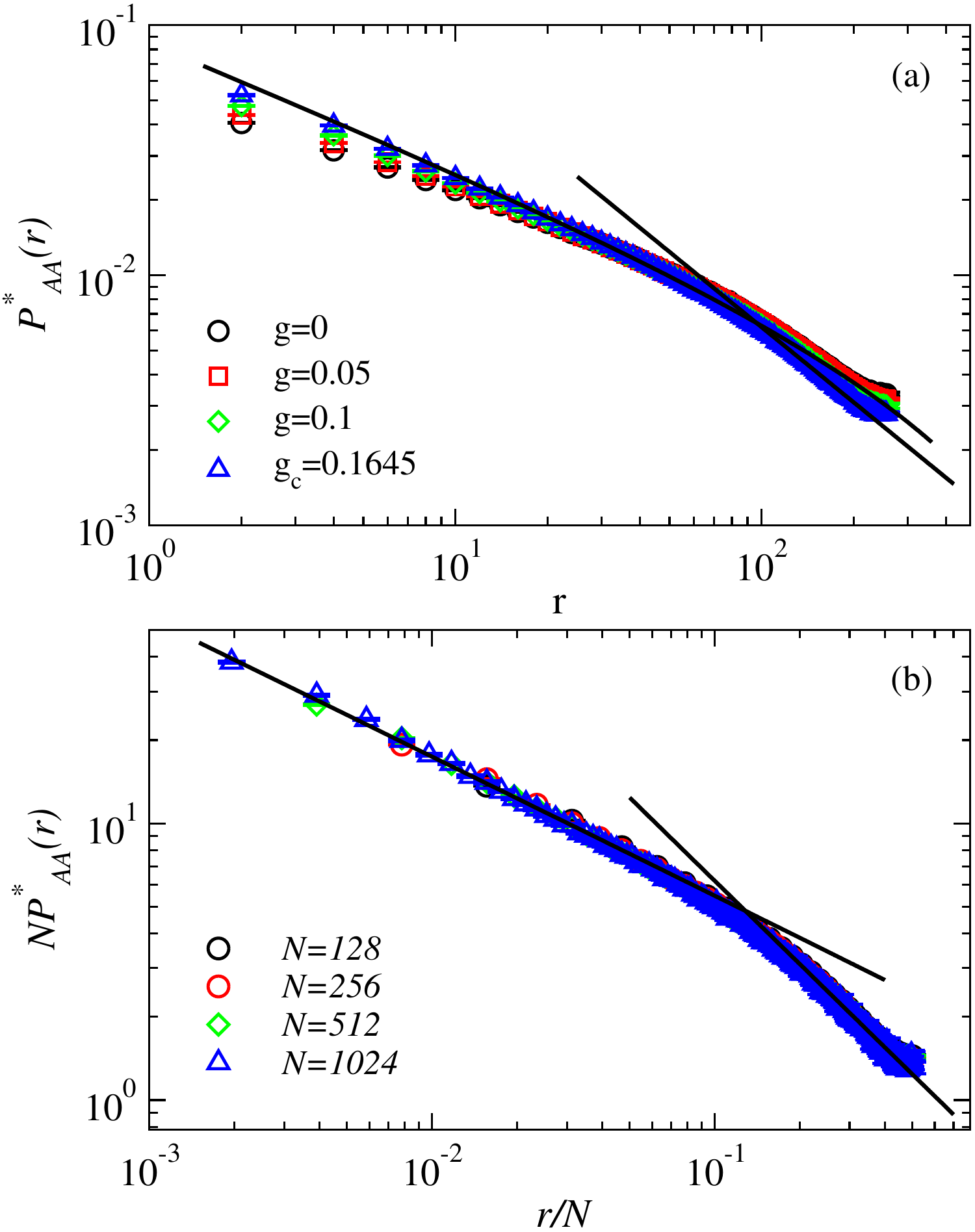}}
\caption{(Color online) Same-sublattice distribution functions for $S=1$ states in the critical phase. (a) Shows results for different coupling ratios for
fixed system size $N=512$, while in (b) results at $g_c$ are re-scaled to achieve data collapse for several system sizes. The lines have slope $1/2$ and $1$
for small and large $r$, respectively.}
\label{figa4}
\end{figure}

As shown in Fig.~\ref{figa4}, the $S=1$ function $P^*_{AA}(r)$ inside the critical phase exhibits an interesting cross-over behavior, most clearly visible at $g=g_c$. The behavior at short distances is well described by the same $r^{-1/2}$ behavior as the corresponding single-spinon function in Fig.~\ref{figa5}. However, at larger distances the behavior changes to $\propto 1/r$. We do not have any explanation for this behavior and it would be interesting to investigate it within bosonization.

\subsection{Break-down of spinons as quasi-particles of a N\'eel state in one dimension}
\label{sec:neel}

\begin{figure}
\centerline{\includegraphics[angle=0,width=8.25cm, clip]{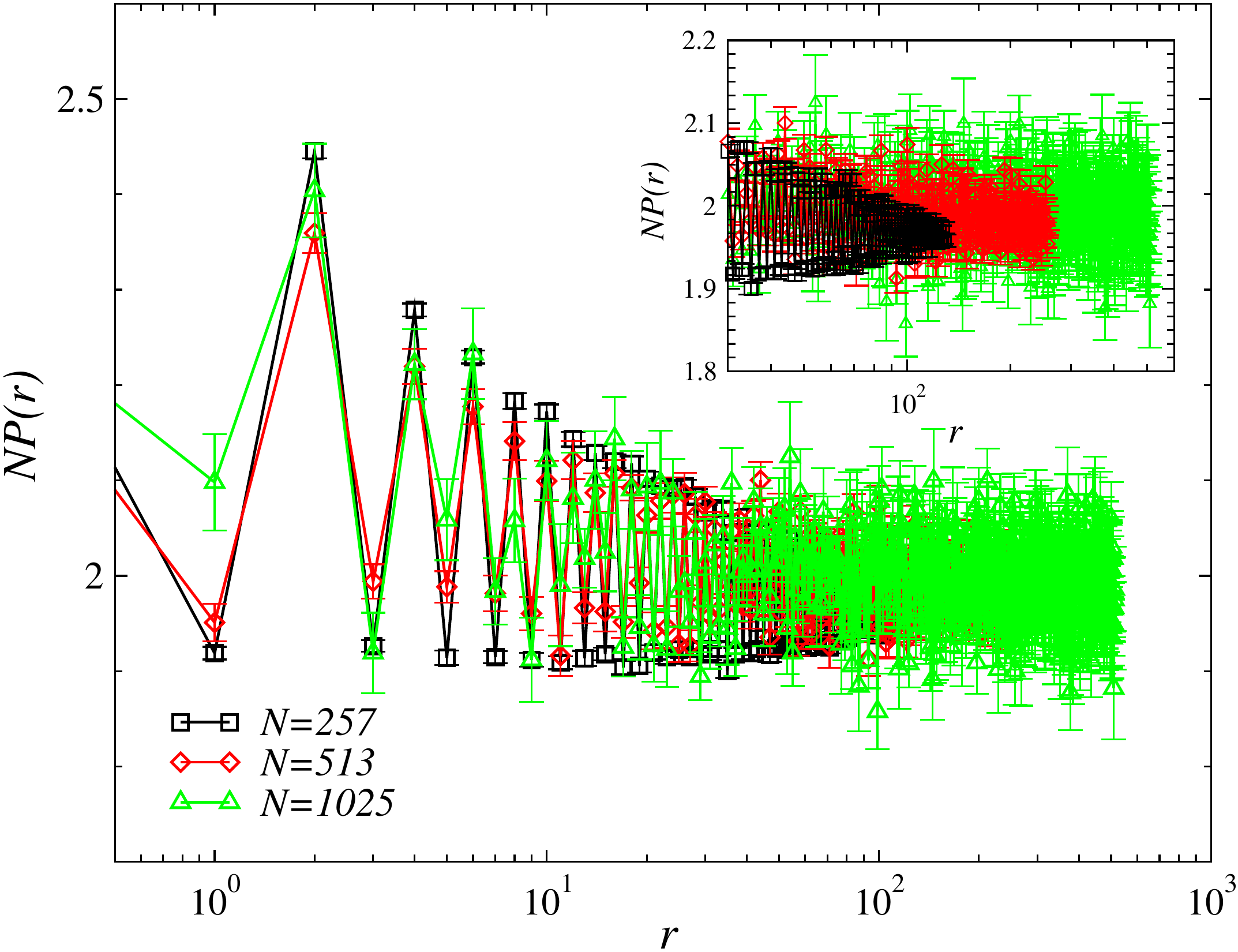}}
\caption{(Color online) Size-scaled spinon overlap function in a N\'eel-ordered chain with total $S=1/2$, computed for chain lengths $N=257,513$, and $1025$.
The asymptotically flat (with even and odd-$r$ branches) distribution  shows that the spinon is not a well-defined quasi-particle in the N\'eel state, as expected. 
The inset shows the tail of the spinon overlap function of N\'eel-ordered chains with a clearer view of the $N=217$ and $513$ data.}
\label{figa2}
\end{figure}

In a long-range ordered N\'eel AFM state, the elementary excitations are spin waves (magnons) carrying spin $S=1$. It is then interesting to ask how the change
in the nature of the excitations is manifested in our spinon distribution functions if the system can be driven to a N\'eel state. The continuous spin-rotational 
symmetry of the ground state of the Heisenberg or J-Q chains cannot be spontaneously broken, however, according to the Mermin-Wagner theorem \cite{Mermin66}.
We can circumvent this limitation on 1D ground states by including long-range interactions, in  which case the theorem does not apply. We here consider unfrustrated
power-law decaying interactions defined by the Hamiltonian
\begin{equation}
H = \sum_{i=1}^{N} \sum_{{\rm odd}~r}^{N/2} J_r {\bf S}_i \cdot {\bf S}_{i+r}, J_r > 0,
\label{eq:neel}
\end{equation}
where there are no couplings for even separations of spins, while for odd separations the coupling is $J_r = 1/r^{\alpha}$. A similar Hamiltonian was studied before 
in Ref.~\onlinecite{Laflorencie05}, where it was found that by tuning the decay exponent $\alpha$ the system undergoes a continuous phase transition from critical states 
when $\alpha > \alpha_c$ to a long-range ordered N\'eel states when  $\alpha < \alpha_c$. The critical power depends on details, e.g., on the strength of the nearest-neighbor 
coupling, and in the cases studied in Ref.~\onlinecite{Laflorencie05} $\alpha_c \approx 2.2$. In Ref.~\onlinecite{Sandvik10e} frustration was added to the model in order to 
drive it to a VBS phase. In our study we are just interested in studying an example of a 1D N\'eel state and choose $J_r = r^{-3/2}$ (odd $r$) in Eq.~(\ref{eq:neel}), for 
which we verified that indeed the system is AFM ordered.

We investigate the single-spinon distribution function $P_{AA}(r)$ in an $S=1/2$ state for odd $N$. In Fig.~\ref{figa2}, we plot $P_{AA}(r)$ scaled by $N$ versus $r$
for different system sizes and find good convergence as a function of the system sizes, although the error bars are large at the largest distances. The behavior here is 
quite different from the previous cases, Figs.~\ref{figa1} and \ref{figa5}, with (i) no vanishing of the probability of odd-$r$ separation and (ii) no decay of the 
rescaled function. The  latter behavior indicates that the spinon here is not a well-defined particle, with no concentration of the net magnetization to within an intrinsic 
wave packet. This is of course not surprising, in the sense that spinons are not expected to be the elementary quasi-particle excitations of the N\'eel state. We had also 
already found above that in the critical state the quasi-particles are only marginal, characterized by power-law overlaps (and hence any further enhancement of 
antiferromagnetic correlations should completely destroy the spinons). It is still interesting to see that the break-down of the spinons as quasi-particles can be 
explicitly observed in the distribution function $P_{AA}(r)$.

\section{Spinon confinement arising from modulated couplings}
\label{sec:j1j2q3}

In order to observe confinement of spinons, we here use a generalized version of the $J$-$Q_3$ model with different nearest-neighbor coupling constants on even and odd 
bonds. The Hamiltonian is
\begin{eqnarray}
&&H=-\sum_{\rm even~i} (J_1C_{i,i+1} + J_2 C_{i+1,i+2})~~~~ \nonumber \\
&&~~~~ -Q_3\sum_{i}C_{i,i+1}C_{i+1,i+2}C_{i+2,i+3}.
\label{j1j2}
\end{eqnarray}
When the modulation parameter $\rho=J_2/J_1 \not=1$, the Hamiltonian itself breaks translational invariance and there is no longer a VBS phase transition with 
spontaneously broken symmetry. If we start in a spontaneously formed VBS $(Q_3/J_1 > g_c$) for $\rho=1$, the ground state is doubly degenerate, but once $\rho > 1$ 
the degeneracy is broken and the ground state is unique. This is expected to confine the spinons, as the string of out-of-phase bonds formed between two separated 
spinons is now associated with an energy cost increasing linearly with the separation, instead of the energy only being associated with the domain walls when $\rho=1$. 
This model was also studied in the presence of an impurity in Ref.~\onlinecite{doretto09}, and it was found that the localization length of the magnetization distribution 
forming around the impurity could be tuned by $\rho$. It was argued that two regions of confinement could be defined; (i) strong confinement, where the size
of the bound state is similar to the standard spin correlation length, and (ii) weak deconfinement, where the bound state is much larger than the correlation
length. Here we find similar behavior for two spinons binding to each other instead of a static impurity. {\it A priori} it is not clear that the situations
are identical, as the impurity-spinon and spinon-spinon potentials are not identical (since a dynamic spinon perturbs its singlet environment differently than
a static impurity).

\begin{figure}
\centerline{\includegraphics[angle=0,width=8cm, clip]{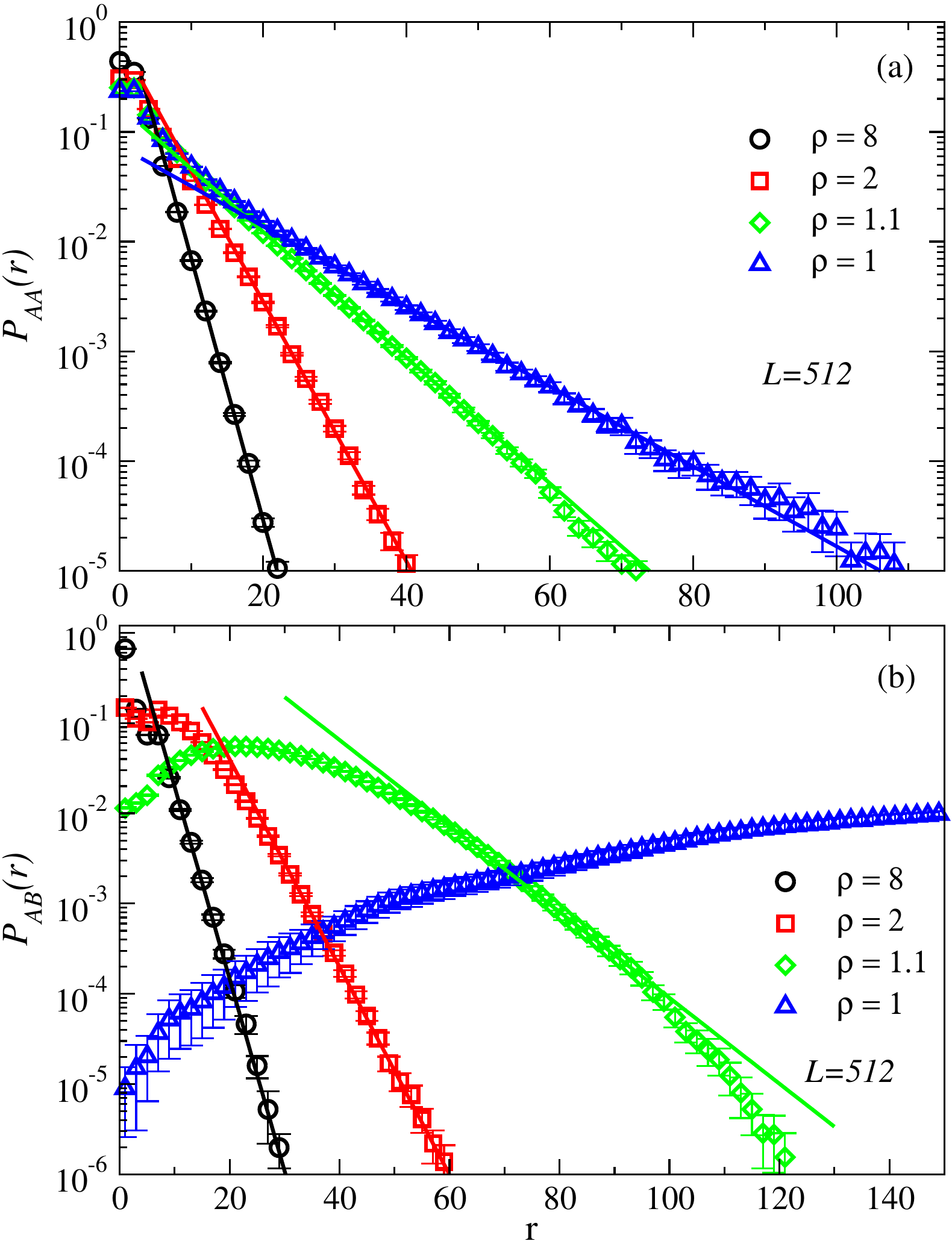}}
\caption{(Color online) Spinon distribution functions in the $J_1$-$J_2$-$Q_3$ chain with $Q_3/J_1=4$ and several values of the modulation parameter $\rho=J_2/J_1$. (a) Shows exponential decays, $P_{AA}(r) \sim e^{-r/\lambda}$, of the single-spinon distribution function of the $S=1/2$ state, demonstrating well-defined spinons with finite intrinsic size $\lambda$. In (b), spinon confinement for $\rho \not =1$ is demonstrated in the spinon-distance distribution function; $P_{AB}(r) \sim e^{-r/\Lambda}$. The size of the bound state (the confinement length scale) decreases as the coupling modulation is increased. Data for $\rho=1$ are graphed for comparison; in this case, the spinons are deconfined and the distribution function does not decay with the separation.}
\label{figb1}
\end{figure}

\begin{figure}
\centerline{\includegraphics[angle=0,width=8cm, clip]{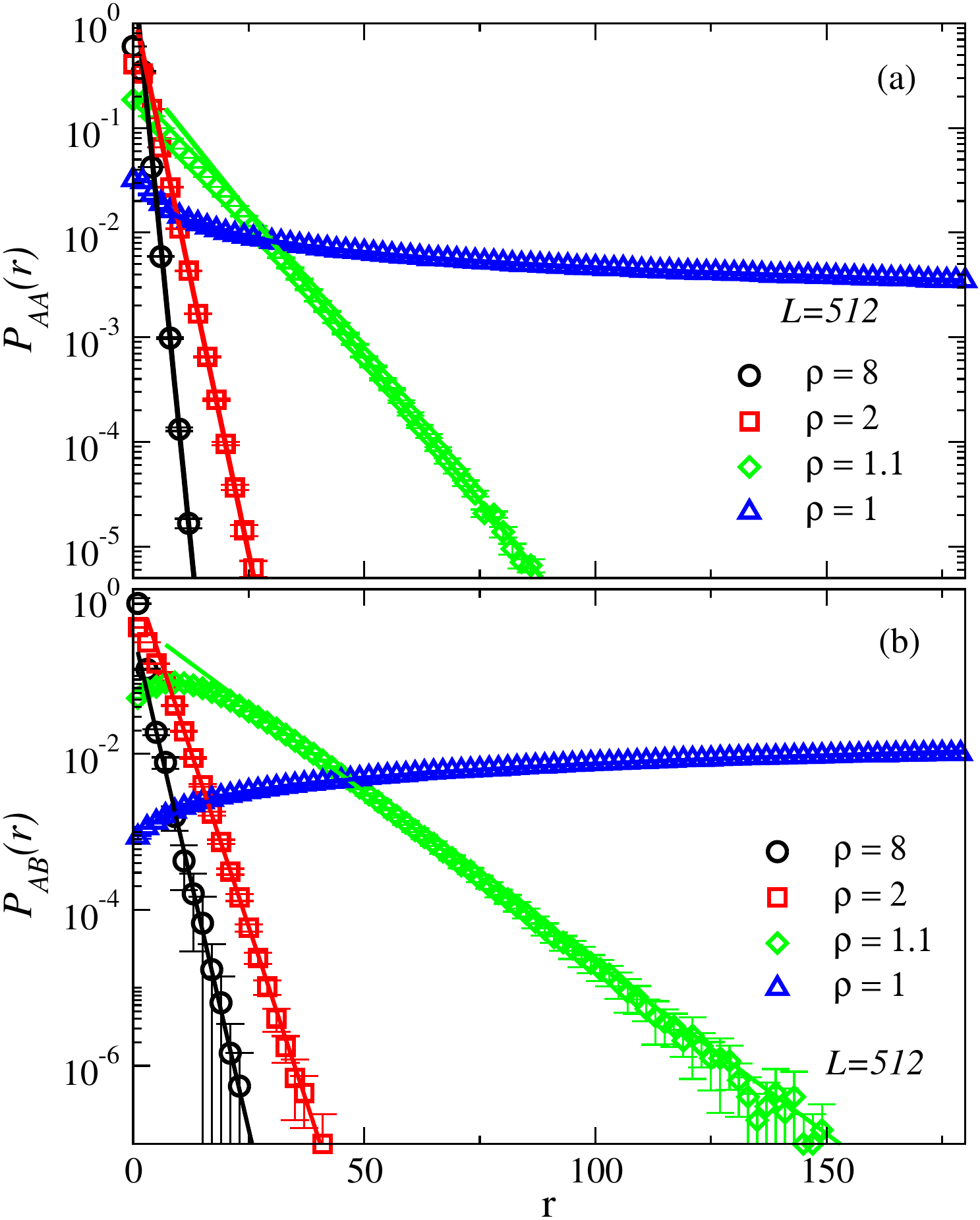}}
\caption{(Color online) The same quantities as in Fig.~\ref{figb1} but with the ratio $Q_3/J_1=g_c=0.1645$. Here, the tuning of the modulation parameter
$\rho$ toward $1$ corresponds to approaching a critical point.}
\label{figb2}
\end{figure}

We first test for confinement deep inside the VBS phase at $g=Q_3/J_1=4$. As shown in Fig.~\ref{figb1}(a), the spinon size $\lambda$ computed from $P_{AA}(r)$ in the $S=1/2$
ground state becomes smaller when the confining potential increases (tuning $\rho$ from $1$ to $8$). Figure ~\ref{figb1}(b) shows that the confinement length 
$\Lambda$ indeed becomes finite once we tune $\rho$ off $1$. For $\rho$ very close to $1$ it is difficult to extract $\Lambda$ because we also need to satisfy
$L \gg \Lambda$ and the calculations become very demanding. Upon increasing $\rho$ we find that $\Lambda$ approaches $\lambda$. 

An interesting observation in Fig.~\ref{figb1}(b) is the maximum developing in $P_{AB}(r)$, seen around $r=20$ for $\rho=1.1$ and moving to $R=N/2$ at the uniform point $\rho=1$. In Sec.~\ref{sec:1D}, we already argued that there is an effective short-range repulsive interaction between the spinons in the uniform chains, and it is natural that these interactions should persist also for some range of $\rho$ away from $1$, although there is also an attractive part binding
the spinons. Thus, we arrive at the conclusion that when $\rho$ is close to $1$ there is a short-range repulsion followed by the linear confining attractive potential 
at longer distances. Judging from the fact that the maximum probability moves toward $r=0$ for larger modulation parameters, $\rho=2,8$ in Fig.~\ref{figb1}(b), 
the role of the short-range repulsion diminishes (leading to the spinon core being ``crushed'')  relative to the linear attractive confinement potential, 
which grows with $\rho$. The cases of $\lambda \approx \Lambda$ and maximum probability at $r=0$ seems very similar to the case of ``strong confinement''
by an impurity in Ref.~\onlinecite{doretto09}, while the case of remaining effects of repulsions pushing the maximum probability away from $r=0$ is
like the ``weak confinement'' case. It would be interesting to compare the two cases more quantitatively, but we leave this for future studies.

We also observe similar behaviors in the dimerized model at the critical $Q_3/J_1$ value, as shown in Fig.~\ref{figb2}. The main difference is that now the
spinon size $\lambda$ diverges as $\rho \to 1$, instead of tending to a finite value in the VBS phase. Both length scales are actually smaller than in the VBS
phase for larger $\rho$, e.g., for $\rho=2$, $\Lambda \approx 2.42(1)$ at $g_c$ while $\Lambda \approx 3.78(4)$ at $g=4$. This implies that the imposed dimerization 
in the critical region has a stronger effect than in the ordered VBS phase. In the critical region all lengths diverge, and, therefore, once we add the explicit
dimerization $\rho\not=1$ it dominates the physics immediately. In contrast, in the VBS phase there are competition effects between the spontaneous VBS and the explicit
dimerization, which apparently reduce the effects on the spinon size and confinement length. Also here we can see a maximum in $P_{AB}(r)$ away from $r=0$,
and $\Lambda$ here is somewhat larger than $\lambda$. It would be interesting to study in detail the divergence of these lengths as $\rho \to 1$ 
and compare them with both the spin and VBS correlation lengths (and also to compare with the impurity-binding case), but we also have to leave this for 
future studies.

\section{HEISENBERG LADDERS}
\label{sec:ladder}

Another way to confine the spinons of the Heisenberg chain is to couple two chains into a ladder, described by the Hamiltonian
\begin{equation}
H =  J_1 \sum_{i=1}^L({\bf S}_i^1\cdot {\bf S}_{i+1}^1 + {\bf S}_i^2\cdot {\bf S}_{i+1}^2)+ J_2\sum_{i=1}{\bf S}_i^1\cdot {\bf S}_i^2,
\label{eq:ladder}
\end{equation} 
where the superscripts $1$ and $2$ label the two chains, $J_1$ is the nearest-neighbor coupling within the chains, and $J_2$ is the inter-chain (rung) 
coupling. It is known that any inter-chain coupling $J_2$ opens a gap in the excitation spectrum and changes the critical correlations to an exponentially decaying
form \cite{Barnes93}. This is true for ladders with any even number of legs, while odd-leg ladders are critical and exhibit the universality of the single 
chain \cite{Dagotto96}. The situation here is similar to single chains of Heisenberg-coupled integer or half-odd-integer spins, with the former always being gapped according 
to the now well confirmed ``Haldane conjecture'' \cite{Haldane83}. The integer-$S$ chains have localized spinons at the ends of open chains, and this is also the case 
(perhaps less surprisingly) in open ladders where a spin is removed from each end. We here investigate the spinon confinement mechanism in the periodic,
translationally invariant ladder.

Gapped triplons ($S=1$), which are the low-lying excitations of ladder systems, have already been observed in the excitation spectrum of real materials by inelastic 
neutron scattering \cite{Thielemann09}. It has been argued that this observation makes the ladder system the simplest condensed matter system where one can in practice
realize a phenomenon similar to quark confinement in particle physics \cite{Lake10}. The energy gap, spin-triplet dispersion relation and the dynamic spin structural 
factor of the Heisenberg two-leg ladder model have also been extensively studied by numerical methods \cite{Barnes93}. 

\begin{figure}
\centerline{\includegraphics[angle=0,width=8cm, clip]{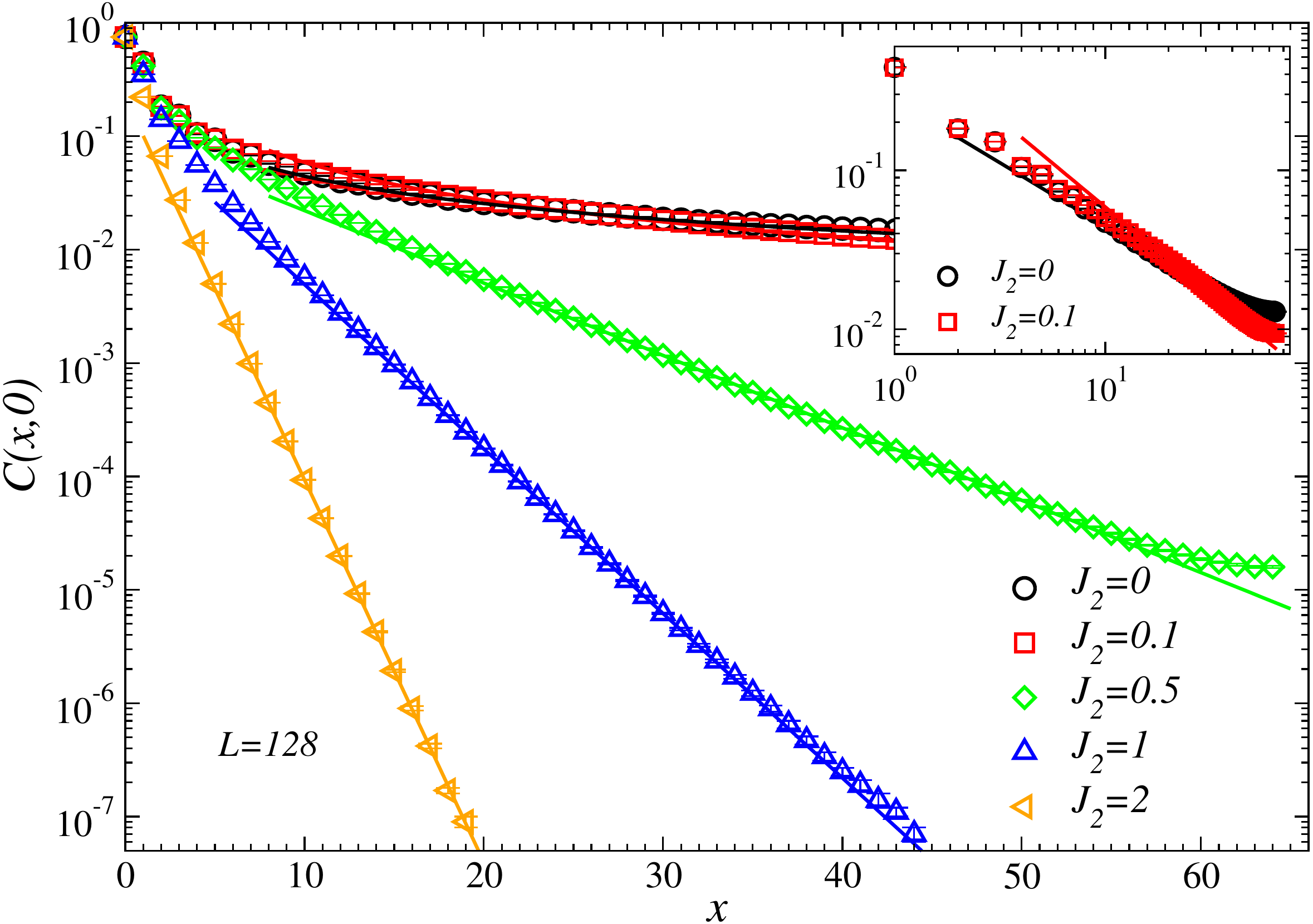}}
\caption{(Color online) Spin correlation function in Heisenberg ladder systems. Here the intra-chain coupling $J_1=1$ and results are shown for several
values of inter-chain couplings $J_2$. $C(x,0)$ decays exponentially when $J_2 \neq 0$ and exhibits the power-law decay of the isolated chain when $J_2=0$. 
In the inset, the correlations are large distances on a log-log scale at $J_2=0$ and  $0.1$. Because here the system length $L$ is smaller than the correlation
length it is not yet possible to observe the exponential decay.}
\label{figc1}
\end{figure}

We begin by discussing the standard spin-spin correlation function in the $S=0$ ground state. We fit it to the form $C(r) \propto e^{-\Delta/\xi}$ when 
$g=J_2/J_1 > 0$, and will later compare the spinon-related length-scales with the correlation length $\xi$. Results are shown in Fig.~\ref{figc1}. 
Note that it is very difficult to extract $\xi$ when $g$ is small, as $\xi$ then becomes large and the system size has to be even larger, $L \gg \xi$. The
inset of Fig.~\ref{figc1} illustrates this problem for $g=0.1$. We here focus on rung couplings sufficiently large for extracting $\xi$ reliably based on
our available ladder sizes.

\begin{figure}
\centerline{\includegraphics[angle=0,width=8cm, clip]{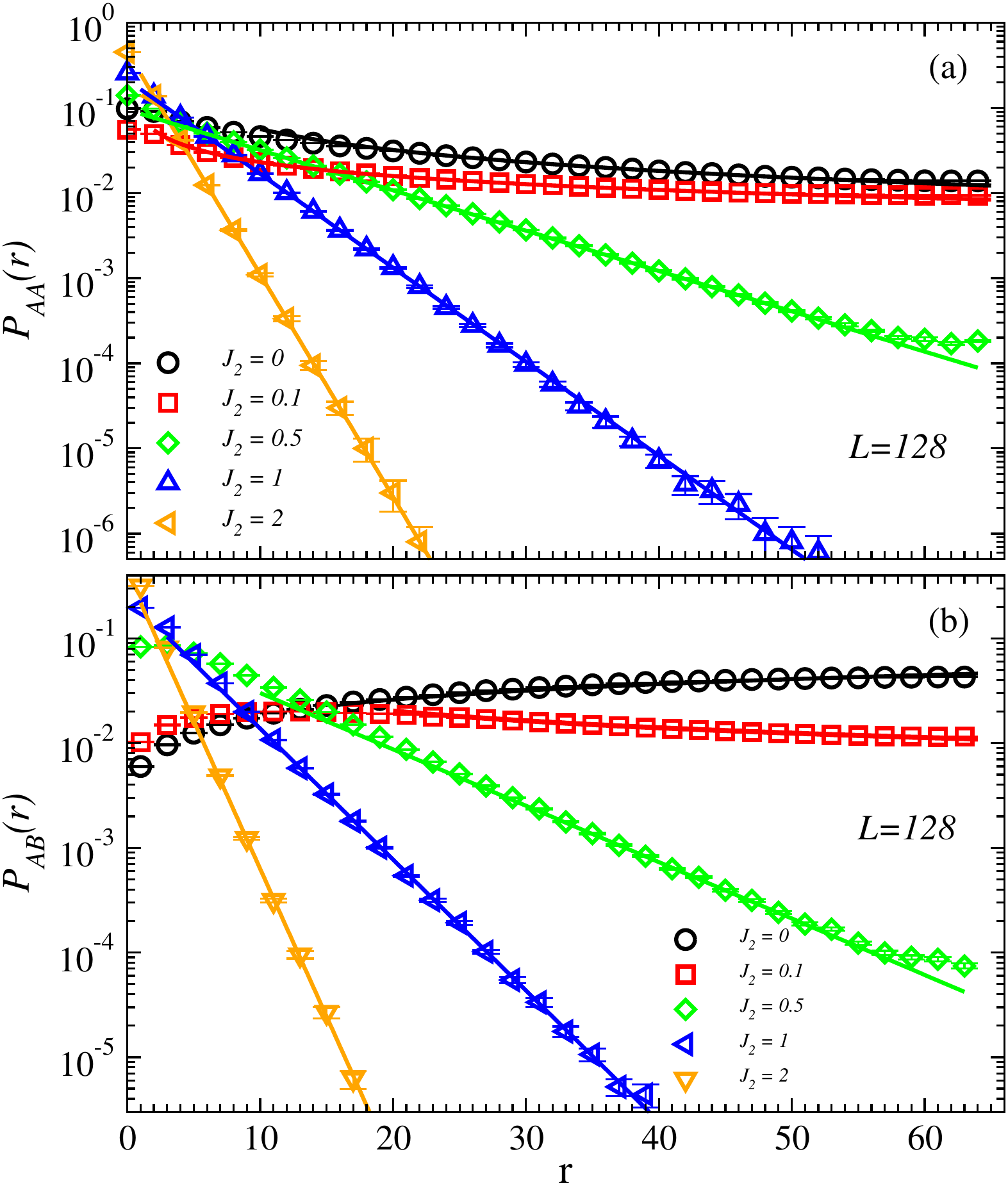}}
\caption{(Color online) Spinon distribution functions in $S=1$ states of Heisenberg ladders with different rung couplings $J_2$. Both distribution functions are
exponentially decaying for $J_2=0.5,1$, and $2$, while for $J_2=0.1$ the system size is not sufficiently large for observing the expected asymptotic exponential decay. }
\label{figc2}
\end{figure}

We now turn to the characterization of the spinons. In the two-leg ladder it is not possible to study a system with an odd number of spins $N$ ($N=2L$) without breaking the translational symmetry of the system (which is a much more severe issue than the boundary subtleties in the single chain, discussed in Sec.~\ref{sec:vbqmc1},
which do not ruin the translational symmetry). 
We here only discuss calculations in the $S=1$ state for even $N$ and present results for the distributions $P^*_{AA}(r)$ and $P_{AB}(r)$ in Fig.~\ref{figc2}. 
As we discussed in Sec.~\ref{sec:1D}, $P^*_{AA}(r)$ can reliably give the intrinsic spinon size $\lambda$ if this length-scale is smaller than the size $\Lambda$ 
of the bound state---in principle one would expect to need $\Lambda \gg \lambda$ but in practice, as shown in Figs.~\ref{figpaaplus} and \ref{figb1}, it seems to 
work also otherwise. In the ladder, the length $\lambda^*$ as extracted from $P^*_{AA}(r)$ is always very similar to $\Lambda$ from $P_{AB}(r)$, however, and, 
therefore, it is not clear whether $\lambda^*$ can be interpreted strictly as the size of an individual spinon, although based on the previous comparisons
one may well argue that it is the case. In the ladder systems, $\lambda^*$ is even somewhat larger than $\Lambda$ , e.g., at $J_{2}=1$, $\lambda^* \approx 3.9$ 
and $\Lambda=3.5$. 

We recently studied a 2D $J$-$Q_3$ model with a VBS state \cite{Tang13}. In that case, an individual spinon in an $S=1/2$ state can be studied and we found that the so
extracted $\lambda$ is considerably smaller than the bound state of two spinons. We interpreted this as being due to a softness of the extended spinons, which
are expected to be a kind of vortices in 2D. Such soft spinons shrink when they are subject to mutual attractive interactions and form a pair. Also there the 
single-spinon length $\lambda^*$ extracted from the $S=1$ state is somewhat larger than $\Lambda$. Given this similarity, we also interpret $\lambda^* \approx \Lambda$
in the Heisenberg ladder as due to softness of the spinons.

\section{Detecting spinons in spin correlations}
\label{sec:crr}

\begin{figure}
\centerline{\includegraphics[angle=0,width=8cm, clip]{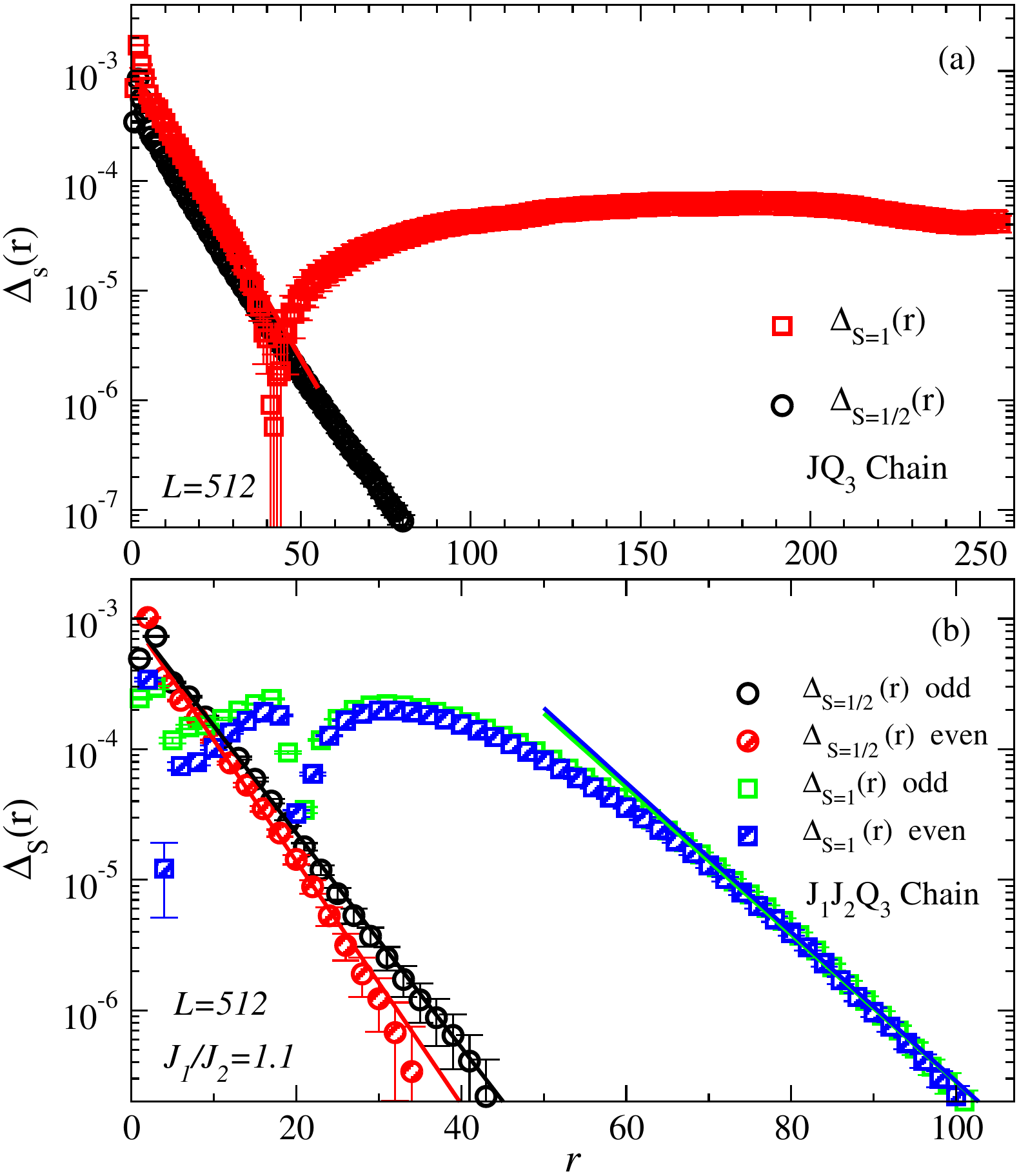}}
\caption{(Color online) Absolute value of the spin correlation function in the $S=1/2$ and $1$ state, after subtraction of the $S=0$ correlation function
according to Eq.~(\ref{deltac}). (a) is for a $J$-$Q_3$ chain with $Q_3/J=4$ and (b) is for a $J_1$-$J_2$-$Q_3$ chain with $J_2/J_1=1.1$ and $Q_3/J_1=4$. In both cases 
the chain length is $L=512$. The sharp dips where the relative errors are large for the $S=1$ quantities correspond to phase shifts [in $S_1(r)$ and $\Delta_1(r)$]. 
In (b) the even-$r$ and odd-$r$ branches are graphed in different colors to show the even-odd effects, while in (a) these effects are too small to be visible. 
All lines correspond to exponential fits.}
\label{figb3}
\end{figure}

The definitions $\Lambda$ and $\lambda$ of the spinon length-scales are closely tied to the VB basis, and the underlying distribution functions are not directly
physically measurable quantities. It is therefore interesting to investigate whether the same length scales also appear in {\it bona fide} quantum-mechanical
expectation values as well. The natural candidate is the standard spin correlation function using the operator (\ref{ssop}) in the total-spin sectors with $S=1/2$ and $S=1$.
It is clear that these correlations overall should not differ significantly from those in the ground state with $S=0$ and we therefore look explicitly at at the 
difference between the two correlation functions, defining
\begin{equation}
\Delta_{S}(r) = C_S(r)-C_0(r),
\label{deltac}
\end{equation}
where the subscript in $C_S$ indicates the spin sector in which the correlations are computed. 
We plot the absolute value of these functions for a $J$-$Q_3$ chain in Fig.~\ref{figb3}(a) 
and for a $J_1$-$J_2$-$Q_3$ chain with a small modulation parameter $\rho=1.1$ in Fig.~\ref{figb3}(b). In both cases, $Q_3$ is relatively large, so that the uniform
$J$-$Q_3$ chain is deep inside the VBS phase.

For $S=1/2$, we find an almost pure exponential decay in Fig.~\ref{figb3}(a), with a decay constant almost the same as the 
single-spinon size $\lambda$ obtained previously 
for this VBS state. As shown in Fig.~\ref{exp3}, $\lambda > \xi_c$, and, thus, the excess correlations in the $S=1/2$ state decay slower than 
those in the $S=0$ state and it is natural to associate these correlations with the intrinsic spinon size. We conclude that  $\lambda$ is 
an actual physical characteristic of the $S=1/2$ state, observable in the long-distance decay of $\Delta_{1/2}(r)$.

In the $S=1$ state, we find an 
interesting structure, where at short distances the behavior follows closely the same exponential decay as in the $S=1/2$ state, while for larger distances there is a 
rather dramatic change, with a phase shift in the staggered correlations (which here is not seen directly as we are graphing only the absolute value, but the shift is 
reflected indirectly in the sharp dip to very small value within a narrow $r$-range), followed by a flattening out of the correlations. The phase shift and subsequent 
flattening out can be understood in terms of deconfined spinons in the following way: Since we are looking at a state with total $S^z=S=1$, the spin correlations at long 
distances are completely dominated by the contributions from the unpaired spins and their transition-graph strings (the singlet background, corresponding to the
loops in the transition graphs, having exponentially decaying correlations). Since these spinons are fixed in the ``up'' state and always reside on different sublattices, 
we will get positive (negative) contributions from odd (even) distances, in contrast to the normal phase of the correlations an antiferromagnet, which is negative (positive)
at odd (even) distances. We find the standard phase of the correlations in the $S=1$ state as well at short distances. Given this, there must be a phase shift at some 
distance $r$. The exact location of the phase shift depends on the model parameters and the chain length in a way which we have not yet disentangled.

\begin{figure}
\centerline{\includegraphics[angle=0,width=8cm, clip]{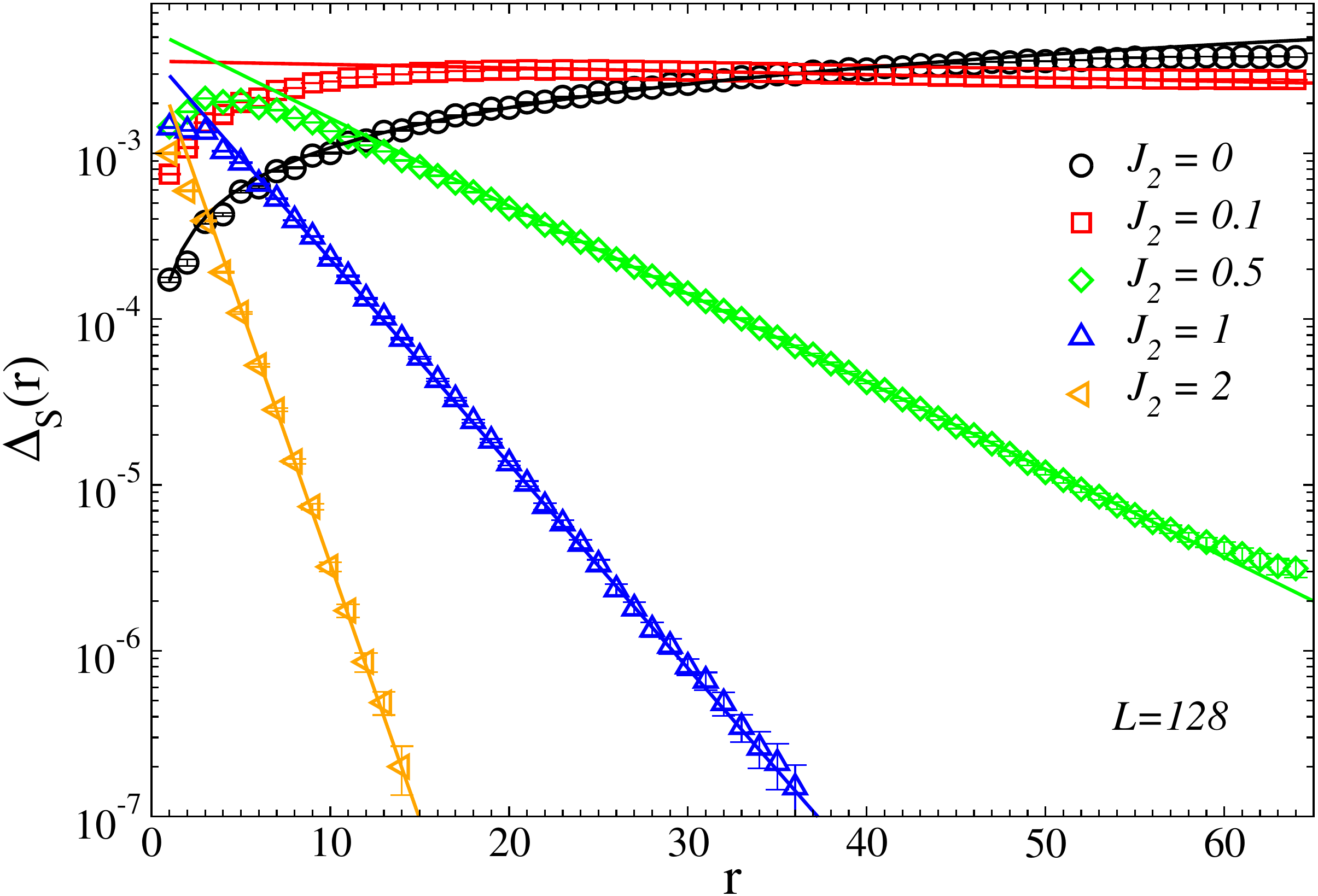}}
\caption{(Color online) Spin correlation difference for Heisenberg ladder systems in the $S=1$ sector. The lines show exponential fits.}
\label{figc3}
\end{figure}

As shown in Fig.~\ref{figc3}, in the case of the ladder systems we do not find any phase shifts and in all cases studied the correlation difference between 
the $S=1$ and $0$ is essentially a pure exponential form. In the ladder we have not found any case where $\Lambda$ is significantly larger than $\lambda$ and
most likely these quantities both diverge in the same way as $J_2/J_1\to 1$. There is therefore no clear regime of weak deconfinement, although the term may
be misleading when the length scales both do become large. We therefore suggest the term {\it marginal deconfinement} to describe this scenario.

\section{SUMMARY AND DISCUSSION}
\label{sec:conclusion}

We have used a computational technique based on valence-bond projector QMC simulations to study the spinon size $\lambda$ and the confinement length $\Lambda$ in 1D 
spin systems. We found that when a system has only one unpaired spinon, the overlap between valence-bond states with unpaired $S^z=1/2$ moment residing at distance $r$ away 
from each other decays as $e^{-r/\lambda}$ in a gapped VBS, where we interpret $\lambda$ as characterizing the intrinsic spinon size. In a critical state, the
overlap instead decays as $r^{-1/2}$, which we interpret as spinons that are only marginal particles, on the verge of losing their identities as quasi-particles.
When the system has two spinons, the distribution's function for the distance between them decays as $e^{-r/\Lambda}$ if the spinons are confined (which we have studied using a modulated pattern of weak and strong coupling constants, which leads to a linear spinon-binding potential), with $\Lambda$ characterizing the size of the bound state. For deconfined spinons (which we have studied in VBS states and critical states), we found that the distribution function instead exhibits a broad peak 
at the largest separation, demonstrating a weak repulsive potential between the spinons. We studied the Heisenberg two-leg ladder system. By tuning the rung coupling,
the system can be driven from a deconfining phase (two decoupled chains) to a confining phase. In this case the spinon size is always similar to the size of the
bound state.

In the Bethe-ansatz solution of the Heisenberg chain, spinons are non-interacting particles (kinks and anti-kinks), but it should be noted that these particles are 
obtained from the original spin degrees of freedom using a highly non-local transformation. What we have probed here is instead more direct measures of the spatial 
``concentration'', $P_{AA}(r)$, of the total magnetization of a single spinon, and the correlations between (essentially) the center of mass of two such distributions, 
$P_{AB}(r)$. Since our calculation projects out the lowest state with given total spin, in the case of $S=1$ the total momentum $k=\pi$ (in the case of a chain
with $N=4n$ sites). Therefore, the spinons here are not propagating, having individual spin $0$ and $\pi$ (these giving the lowest possible energies in light of 
the des Clauseaux-Pearson dispersion). In principle our calculations can also handle total momentum away from $k=\pi$, but in practice, due to phase problems in the Monte Carlo sampling, we are restricted to momenta close to $0$ and $\pi$.

In the future, it would be interesting to more exhaustively characterize all the length scales of the system (including $\lambda$, $\Lambda$, as well as the 
spin and VBS correlation lengths) and their divergences under the various conditions afforded by the models we have performed initial studies on here.

\section*{ACKNOWLEDGEMENT}
This work was supported by the NSF under Grants No. DMR-1104708 and DMR-1410126. ----------------------------     

\null

\end{document}